\documentclass[11pt,letterpaper,oneside]{article}
\usepackage[margin=1in]{geometry}
\usepackage[T1]{fontenc}
\usepackage{lmodern}
\usepackage{microtype}
\usepackage{amsmath,amssymb,mathtools,amsthm}
\usepackage{stmaryrd}
\SetSymbolFont{stmry}{bold}{U}{stmry}{m}{n}
\usepackage{booktabs}
\usepackage{enumitem}
\usepackage{xcolor}
\usepackage{bm}
\usepackage{algorithm}
\usepackage{algpseudocode}
\usepackage{comment}
\usepackage[hidelinks]{hyperref}
\usepackage[capitalise,noabbrev]{cleveref}

\makeatletter
\providecommand{\theHALG@line}{}
\renewcommand{\theHALG@line}{\thealgorithm.\arabic{ALG@line}}
\makeatother

\theoremstyle{plain}
\newtheorem{theorem}{Theorem}[section]
\newtheorem{lemma}[theorem]{Lemma}
\newtheorem{proposition}[theorem]{Proposition}
\newtheorem{corollary}[theorem]{Corollary}
\theoremstyle{definition}
\newtheorem{definition}[theorem]{Definition}
\theoremstyle{remark}

\newcommand{\bZ}{\mathbb{Z}}
\newcommand{\bQ}{\mathbb{Q}}
\newcommand{\bR}{\mathbb{R}}
\newcommand{\bC}{\mathbb{C}}
\newcommand{\imag}{\mathsf{i}}

\newcommand{\calR}{\mathcal{R}}
\newcommand{\calK}{\mathcal{K}}
\newcommand{\RealExt}{{\calK}_{\bR}}

\newcommand{\ModMultCost}{\mathsf{M}}
\newcommand{\PolyMultCost}{\mathsf{M}}

\newcommand{\ct}{\mathsf{ct}}
\newcommand{\pt}{\mathsf{pt}}

\newcommand{\sk}{\mathsf{sk}}
\newcommand{\swk}{\mathsf{swk}}

\newcommand{\Gal}{\mathsf{Gal}}

\newcommand{\Enc}{\ensuremath{\mathsf{Enc}}}
\newcommand{\Dec}{\ensuremath{\mathsf{Dec}}}
\newcommand{\Ecd}{\ensuremath{\mathsf{Ecd}}}
\newcommand{\Dcd}{\ensuremath{\mathsf{Dcd}}}

\newcommand{\RS}{\ensuremath{\mathsf{RS}}}
\newcommand{\KS}{\ensuremath{\mathsf{KS}}}

\newcommand{\Add}{\ensuremath{\mathsf{Add}}}
\newcommand{\Mult}{\ensuremath{\mathsf{Mult}}}
\newcommand{\PtMult}{\ensuremath{\mathsf{PtMult}}}
\newcommand{\Aut}{\ensuremath{\mathsf{Aut}}}
\newcommand{\Rot}{\ensuremath{\mathsf{Rot}}}

\newcommand{\BTS}{\ensuremath{\mathsf{BTS}}}

\newcommand{\StC}{\ensuremath{\mathsf{S2C}}}
\newcommand{\CtS}{\ensuremath{\mathsf{C2S}}}
\newcommand{\EvalMod}{\ensuremath{\mathsf{EvalMod}}}
\newcommand{\ModRaise}{\ensuremath{\mathsf{ModRaise}}}

\newcommand{\canorm}[1]{\left\lVert {#1} \right\rVert_\infty^\mathsf{can}}

\DeclarePairedDelimiter{\abs}{\lvert}{\rvert}
\DeclarePairedDelimiter{\norm}{\lVert}{\rVert}
\DeclarePairedDelimiter{\round}{\lfloor}{\rceil}
\DeclarePairedDelimiter{\coeffvec}{\llbracket}{\rrbracket}   %

\hypersetup{
  pdftitle={Encrypted Fixed-Point Multiplication in Quasilinear Complexity},
  pdfauthor={Jung Hee Cheon and Junho Lee}
}

\title{Encrypted Fixed-Point Multiplication in Quasilinear Complexity}
\author{Jung Hee Cheon\textsuperscript{1,2}\qquad Junho Lee\textsuperscript{1}\\[1ex]
\textsuperscript{1}Seoul National University\\
\textsuperscript{2}CryptoLab Inc.\\[1ex]
\texttt{\{jhcheon,abab9579\}@snu.ac.kr}}
\date{}

\begin{document}

\hypersetup{pageanchor=false}
\begin{titlepage}
\maketitle
\thispagestyle{empty}
\begin{abstract}
\normalsize
Homomorphic encryption enables computation on encrypted data,
but its high computational overhead remains a central obstacle.
We overcome this obstacle asymptotically for high-precision fixed-point arithmetic.
In this paper, we introduce focused encoding,
which represents a fixed-point number of $n$-bit precision
using a polynomial of degree $O(n/\log n)$ over a polynomially bounded modulus
by concentrating precision at one embedding.
We then introduce an error-shaping technique for homomorphic operations to preserve the encoding.
This yields an encrypted fixed-point multiplication algorithm for the CKKS scheme
that runs in online time $O(n \log n)$ at $n$-bit precision for any fixed multiplicative depth,
with security parameter $\lambda = O(n/\log^2 n)$.

Further, we develop bootstrapping for focused encoding to overcome the quadratic cost of conventional bootstrapping.
We bootstrap ciphertext coefficients at an enlarged scale
and shape the resulting error to preserve focused encoding,
at the cost of additional modulus consumption.
Then, to reduce the remaining logarithmic overhead,
we decompose the ciphertext into components over a smaller subring and bootstrap each separately.
This reduces the online cost per multiplication, including bootstrapping, to
$O(n \log n + n (\log \lambda)^3 \log \log \lambda)$ for security parameter $\lambda = O(n/\log^3 n)$.

\end{abstract}
\end{titlepage}

\pagenumbering{roman}
\hypersetup{pageanchor=true}
\setcounter{tocdepth}{2}
\tableofcontents
\clearpage

\pagenumbering{arabic}

\section{Introduction}

Encryption protects the confidentiality of data in storage and communication,
but conventional computation requires exposing that data in plaintext,
problematic for sattistical analysis, machine-learning inference, and other computations on confidential data.
Homomorphic encryption allows a server to compute directly on ciphertexts:
The server need not hold the decryption key or receive the intermediate plaintext values.
This makes homomorphic encryption a natural approach to computation on private data,
but its computational overhead has remained a central obstacle.

For numerical computation, this overhead depends on both the security parameter
and the precision of the values being processed.
A basic operation is fixed-point multiplication: given two numbers represented to $n$ bits of precision,
compute their product to the required precision.
In plaintext, this operation reduces to integer multiplication and rounding,
and admits quasilinear bit complexity~\cite{BZ10,HV21}.
It is natural to ask how much additional work is necessary
when the inputs and output must remain encrypted.
We focus on the cost of multiplying a single pair of high-precision numbers,
including the cost of maintaining ciphertexts for subsequent computation.

Homomorphic encryption schemes based on Ring-Learning With Errors (RLWE)~\cite{LPR10} provide efficient arithmetic on encrypted data
by batching multiple slots with SIMD.
BFV~\cite{Bra12,FV12} and BGV~\cite{BGV12} support exact arithmetic over plaintext rings,
while CKKS~\cite{ckks17} supports approximate arithmetic on vectors of real or complex numbers.
Their ability to pack many values into one ciphertext can amortize the cost of a ring operation
across many independent computations.
For a single high-precision value, however, the cost of the entire ciphertext must be accounted for.
In the standard CKKS encoding, $n$ bits of precision require a scale factor larger than $n$ bits,
and multiplication followed by rescaling consumes a comparable number of modulus bits.
The ciphertext modulus must accommodate this scale, while the ring dimension must also grow linearly
to maintain security as the modulus increases~\cite{Albrecht17}.
Consequently, the conventional representation entails a quadratic dependence on input plaintext size,
even before accounting for the cost of arithmetic on ciphertexts.

One can instead represent the inputs as encrypted bits or small digits and evaluate
an arithmetic circuit, as in the gate-bootstrapping approach of TFHE~\cite{CGGI20}.
In particular, Liu and Wang~\cite{LW23BatchII} achieve an amortized bit cost of
$\widetilde{O}(\lambda)$ per gate through batch bootstrapping.
However, applied to a multiplication circuit of size $O(\ModMultCost(n))$,
this gives a $\widetilde{O}(\lambda\ModMultCost(n))$ bound,
where $\ModMultCost(n)$ denotes the cost of plaintext integer multiplication.
Thus, batching alone does not establish a bound matching plaintext multiplication
when the security parameter is considered to be large.

We study this question in a model that permits preprocessing depending on the parameters,
but independent of the inputs.
The preprocessing can be reused across subsequent computations;
the online cost measures the work performed once the inputs are available.
Our central question is:
\begin{quote}
    \itshape
    Can encrypted fixed-point multiplication match the asymptotic cost of plaintext
    multiplication in an appropriate security regime, including the bootstrapping
    needed for continued computation, with fixed input-independent preprocessing?
\end{quote}

\subsection{Contributions}

Our starting point is \emph{focused encoding}, which represents a single number
at $n=\Theta(N\log N)$ bits of precision with a polynomial of degree $N$
with polynomially bounded coefficients.
We employ a focal element,
shaping the error to be exponentially small at the focus embedding and polynomially bounded elsewhere.
As a result, we obtain quasilinear levelled multiplication.
However, conventional bootstrapping goes through an ordinary coefficient encoding at the full message precision,
reintroducing quadratic complexity.

We overcome this bottleneck by bootstrapping the small coefficients of the encoded polynomial,
and then restoring the focused error shape.
To do this, we find a polynomial that is multiple of the focal element that is congruent to one modulo the output modulus,
which is obtained as a product of focal element and its modular inverse.
Multiplying the polynomial to the ciphertext before rescaling,
this preserves the message while giving the error the desired shape.
While this requires additional modulus to absorb the noise amplification,
it is sufficient to yield quasilinear bootstrapping which preserves the message precision.

Then, to reduce the remaining logarithmic overhead,
we perform coefficient bootstrapping over a smaller subring.
As precision grows, the full ring exceeds the ring dimension needed for secure bootstrapping
at the chosen modulus and security parameter $\lambda$.
Hence, we sample the secret key from the subring,
which allows us to split the ciphertext into components and be bootstapped separately.
After recombining the refreshed components,
we use the focal element $u$ to shape their combined error.

Together, these techniques give quasilinear levelled encrypted multiplication
and bootstrapping whose cost is quasilinear in the precision and polylogarithmic in the security parameter.
We allow fixed, input-independent preprocessing,
which may take exponential time and is reusable across computations.

\begin{theorem}[Informal]
  Using an $O(n\log n)$-time integer multiplication algorithm,
  levelled encrypted fixed-point multiplication at $n$-bit precision
  has online time complexity $O(n \log n)$
  for any fixed multiplicative depth,
  provided $\lambda = O(n/\log^2 n)$.
  Including bootstrapping for continued computation,
  the online time per multiplication is
  $O\!\left(n\log n+n(\log\lambda)^3\log\log\lambda\right)$,
  provided $\lambda=O(n/\log^3 n)$.
\end{theorem}

The formal statements are given in \Cref{thm:mult-levelled,thm:bootstrapping-focused,thm:mult-fhe}.

We also give an input encoder of quasilinear time complexity
that passes through intermediate encodings in a tower of cyclotomic rings,
avoiding the quadratic expansion of the direct encoding formula.

\subsection{Related Works}

\subsubsection{High-Precision Homomorphic Arithmetic}
Several works reduce the cost of high-precision arithmetic by changing the plaintext representation.
CLPX~\cite{CLPX18} uses polynomial plaintext moduli to represent large integers with reduced noise growth,
and GBFV~\cite{GV25} extends this approach to support SIMD packing and bootstrapping.
Geelen and Vercauteren~\cite{GV26BTS} improve GBFV bootstrapping by evaluating
SlotToCoeff entirely in GBFV and reducing the work performed in ordinary BFV.
Kim~\cite{Kim26GBFV} instead bootstraps the ciphertext noise through CKKS,
supporting large-precision GBFV instances, including CLPX.
These results concern modular arithmetic with a \emph{limited} choice of modulus.

On the other hand, Gao and Zheng~\cite{GZ26} implement modular arithmetic on top of CKKS
by representing words across multiple CKKS slots.
Wei et al.~\cite{WMGZ26} and Liu et al.~\cite{LMWW26ALU} extend this approach to
large prime moduli and arbitrary plaintext moduli, respectively,
with conversions to Boolean or digit representations using a constant number of bootstrapping calls.
Jang and Lee~\cite{JL26} also support large-characteristic finite-field arithmetic, including modular arithmetic, through bi-spectral encoding.
However, without restrictions on the plaintext modulus, their encoding conversions require dense $n \times n$ matrix,
incurring $\Omega(n^2)$ bit complexity, despite the constant number of bootstrapping calls.

We note that there is no straightforward method to reduce fixed-point arithmetic to modular arithmetic with fixed modulus
without additional operations like bit decomposition. 

For approximate arithmetic, Tuple-CKKS~\cite{CCKS23} decomposes ciphertexts into multiple components,
but there is no known methods for bootstrapping Tuple-CKKS ciphertexts.
High-precision bootstrapping has also been developed through improved modular approximation~\cite{lee2021high},
repeated bootstrapping in Meta-BTS~\cite{BCCKK22},
and discrete-CKKS-based precision enhancement~\cite{CKSS25}.
While these techniques improve the precision of bootstrapping, it does not achieve quasilinear complexity for a multiplication.

We obtain a fixed-point encrypted multiplication algorithm in quasilinear complexity,
and obtain the same complexity as plaintext under certain security conditions.

\subsubsection{Concurrent Work}
Several concurrent works trade slot capacity for precision in approximate homomorphic encryption,
using a public ring element to suppress error on focus slots while allowing larger error in background slots.
Dumezy~\cite{Dumezy26} proposes a framework including bootstrapping,
but their bootstrapping passes through an ordinary coefficient encoding,
whose scale and base modulus must accommodate the full target precision.
Thus, $n$-bit precision still requires $\Omega(n)$ modulus bits at this stage,
giving at least quadratic time complexity.
Ma et al.~\cite{MLWW26} introduce a similar framework named GCKKS, which did not contain bootstrapping.
Their follow-up work~\cite{MLWW26BTS} combines
low-precision overflow extraction with iterative ciphertext splitting and precision doubling.
However, the ciphertext count doubles at each cleaning stage,
costing $O(n)$ homomorphic ring operations to reach $n$-bit precision.
Liang et al.~\cite{LLFD26} extends to different precision requirements across slots, but does not contain bootstrapping.
Meanwhile, our earlier preprint~\cite{CL26} presented the framework as focused encoding and bootstrapping using focal units.

Our present construction is the first to obtain quasilinear complexity in fixed-point multiplication including bootstrapping,
and we achieve this by refreshing the small coefficients over a subring
and then restores the focused error shape, avoiding this expansion into many full-ring ciphertexts.

\section{Preliminaries}

\subsection{Notations}
\label{sec:prelim:notation}

Write $[d]=\{0,\ldots,d-1\}$.
Let $\odot$ denote componentwise multiplication,
and $\bm{e}_r$ the $r$-th standard basis vector, indexed by $r\in[d]$.
$\log$ denotes the logarithm with base $2$.

Let $N$ be a power-of-two integer,
and denote $\calR = \bZ[X]/(X^N+1)$,
$\calK = \bQ[X]/(X^N+1)$,
$\RealExt = \calK \otimes_\bQ \bR = \bR[X]/(X^N+1)$.
The $\ell^\infty$, $\ell^2$, and $\ell^1$-norms on $\RealExt$ are defined with respect to its real coefficient vector.
Given $m \in \RealExt = \bC[X]/(X^{N/2} - \imag)$,
$\coeffvec{m} \in \bC^{N/2}$ denotes its complex coefficient vector.

For $m \in \calK$ and a modulus $Q$, $[m]_Q$ denotes the centered remainder mod $Q$.
For $m\in\RealExt$, $\round{m}\in\calR$ denotes coefficientwise rounding to the nearest integer.
Given an invertible element $u \in \calR_Q$, denote its modular inverse as $[u^{-1}]_Q$.

For a positive integer $M$, write $\zeta_M=\exp(2\pi\imag/M)$, and set $\zeta=\zeta_{2N}$.
For $j\in[N/2]$, $\sigma_j : \RealExt \to \bC$ denotes evaluation at $\zeta^{5^j}$,
and the canonical embedding is $\sigma(m)=(\sigma_j(m))_{j\in[N/2]}\in\bC^{N/2}$.
For $m \in \RealExt$, its canonical norm is $\canorm{m} = \norm{\sigma(m)}_\infty$.
Write $\kappa_j$ for the automorphism $X\mapsto X^{5^j}$.

\subsection{Fixed-Point and Polynomial Multiplication}

We concern ourselves with fixed-point multiplication at $n$-bit precision.
Given $x,y \in [-1,1]$ represented in fixed-point form with $n$ fractional bits,
the task is to compute an approximation $z$ satisfying
\[
    \abs{z-xy} \le 2^{-n}.
\]
Fixed-point multiplication at $n$-bit precision and modular
multiplication with an $n$-bit modulus both have bit complexity
$O(\ModMultCost(n))$, using fast rounding and division~\cite{BZ10}.
Sch\"onhage--Strassen multiplication gives
\[
    \ModMultCost(b)=O(b\log b\log\log b),
\]
while Harvey and van der Hoeven~\cite{HV21} achieve $O(b\log b)$.

For polynomials, consider binary representation of coefficients.
We denote $\PolyMultCost(N,b)$ to be the bit cost of multiplying two polynomials
of degree less than $N$
with coefficients of $b$-bit size.
By Kronecker substitution~\cite{BZ10},
\[
  \PolyMultCost(N,b) = O(\ModMultCost(N (b + \log N))).
\]

Residue Number System (RNS) and Number Theoretic Transform (NTT) representation
can reduce the time complexity for specific cases,
but we do not consider them here.

\subsection{CKKS Scheme}
\label{sec:prelim:ckks}

CKKS~\cite{ckks17} is an RLWE-based levelled homomorphic encryption scheme that supports approximate computation on a vector of complex numbers.
Fix an error distribution $\chi$ on $\calR$ within $\norm{e}_\infty \le \Sigma$,
e.g.\ a discrete Gaussian truncated at $\Sigma$.
It consists of the following algorithms.
\begin{itemize}[leftmargin=*]
    \item $\mathsf{KeyGen}(1^\lambda)$.
        Samples a secret key $\sk \gets \mathcal{HWT}(h)$,
        whose coefficient vector is uniform over $\{0, \pm 1\}^N$ of Hamming weight $h$.
    \item $\Ecd(\bm{z}; \Delta)$. Given $\bm{z} \in \bC^{N/2}$ and a scaling factor $\Delta$,
        returns $\round{\Delta \sigma^{-1}(\bm{z})} \in \calR$.
    \item $\Dcd(\pt; \Delta)$. Given $\pt \in \calR$, returns $\sigma(\pt)/\Delta \in \bC^{N/2}$.
    \item $\Enc_\sk(\pt)$. Given $\pt \in \calR$,
        samples a uniform mask $a \in \calR_Q$ and an error $e_\Enc \gets \chi$,
        and returns $(\pt, 0) + (- a \cdot \sk + e_\Enc, a)$.
    \item $\Dec_\sk(\ct)$. Given $\ct = (b, a) \in \calR_Q^2$,
        returns $[b + a \cdot \sk]_Q$.
    \item $\KS_{\sk \to \sk'}(\ct)$.
        Given $\ct = (b, a) \in \calR_Q^2$ under key $\sk$ and
        a switching key $\swk_{\sk \to \sk'} \in \calR_{PQ}^2$,
        an encryption of $P \cdot \sk$ under $\sk'$ for an auxiliary modulus $P$,
        returns
        \[
            \KS_{\sk \to \sk'}(\ct) = (b, 0) + \round{P^{-1} a \cdot \swk_{\sk \to \sk'}} \in \calR_Q^2.
        \]
    \item $\RS_q(\ct)$.
        Given $\ct = (b, a) \in \calR_{Q}^2$ with $q \mid Q$,
        returns $(\round{q^{-1} b}, \round{q^{-1} a}) \in \calR_{Q/q}^2$.
    \item $\Add(\ct_1, \ct_2)$ returns $\ct_1 + \ct_2 \in \calR_Q^2$.
    \item $\Mult(\ct_1, \ct_2)$.
        Given $\ct_i = (b_i, a_i) \in \calR_{Q}^2$,
        computes $(d_0, d_1, d_2) = (b_1 b_2, a_1 b_2 + a_2 b_1, a_1 a_2)$
        and returns
        \[
            \RS_\Delta((d_0, d_1) + \KS_{\sk^2 \to \sk}((0, d_2))) \in \calR_{Q/\Delta}^2.
        \]
    \item $\Aut_\kappa(\ct)$.
        Given $\ct = (b, a)$ and $\kappa \in \Gal(\calK/\bQ)$,
        returns $\KS_{\kappa(\sk) \to \sk}((\kappa(b), \kappa(a)))$.
\end{itemize}

The following correctness lemmas record worst-case error bounds that are well-established.
\begin{lemma}[Encryption]
    \label{lem:ckks-encryption}
    Given a message $m \in \calR$, its encryption $\ct = \Enc_\sk(m)$ satisfies
    \[
        \Dec_\sk(\ct) = m + e_\Enc \pmod{Q}, \qquad \canorm{e_\Enc} \le E_\Enc := N \Sigma.
    \]
\end{lemma}
\begin{lemma}[Key-switching and rescale]
    \label{lem:ckks-ks-rs}
    Suppose $q \mid Q$ and $P \ge Q$.
    Given a ciphertext $\ct \in \calR_Q^2$,
    \begin{align*}
        \Dec_{\sk'}(\KS_{\sk \to \sk'}(\ct)) &= \Dec_\sk(\ct) + e_\KS \pmod{Q},
        \qquad \canorm{e_\KS} \le E_\KS := \frac{N^2 \Sigma}{2} + \frac{1 + h}{2} N, \\
        \Dec_\sk(\RS_q(\ct)) &= \tfrac{1}{q} \Dec_\sk(\ct) + e_\RS \pmod{Q/q},
        \qquad \canorm{e_\RS} \le E_\RS := \frac{1+h}{2} N.
    \end{align*}
\end{lemma}

We sometimes omit the scale factor in \Ecd{} and \Dcd{} for exposition.

\subsection{CKKS Bootstrapping}
\label{sec:prelim:ckks-bts}

Bootstrapping is a procedure that refreshes a ciphertext whose modulus has been consumed by levelled multiplications,
restoring it to a higher level so that computation can continue.
The contemporary CKKS bootstrapping algorithms are made of the following steps.
\begin{itemize}[leftmargin=*]
    \item \ModRaise.
        Given a ciphertext $\ct \in \calR_{q_0}^2$ encrypting $m \in \calR$ and a higher modulus $Q > q_0$,
        returns its lift $\hat{\ct} \in \calR_Q^2$ encrypting $\hat{m} = m + q_0 I$ for some bounded overflow $I \in \calR$.
    \item \CtS.
        Given a ciphertext encrypting $\hat{m} \in \calR$,
        evaluates inverse DFT to return a ciphertext encrypting $\Ecd(\coeffvec{\hat{m}})$.
    \item \EvalMod.
        Given a ciphertext encrypting $\Ecd(\coeffvec{m} + q \coeffvec{I})$,
        evaluates modular reduction to remove the overflow and return a ciphertext encrypting $\Ecd(\coeffvec{m})$.
    \item \StC.
        Given a ciphertext encrypting $\Ecd(\coeffvec{m})$,
        evaluates forward DFT to return a ciphertext encrypting $m \in \calR$.
\end{itemize}

\subsubsection{Modulus Raising and Homomorphic Modular Reduction}

The \ModRaise{} takes centered remainder of the ciphertext components, so that it encrypts
\[
    [b]_{q_0} + [a]_{q_0} \cdot \sk = m + q_0 I.
\]
It introduces the bounded overflow $q_0 I \in \calR$ with $\norm{I}_\infty \le K$, $K = O(h)$,
and rest of the bootstrapping is dedicated to removing the overflow.

The \EvalMod{} evaluates the modular reduction by $q_0$ to remove the overflow in slots.
Classical constructions evaluate polynomials approximating the modular reduction
using periodic trigonometric functions~\cite{EC:CHKKS18,HK20}.

\subsubsection{Homomorphic Discrete Fourier Transformation}
\label{sec:prelim:dft-decomp}

The \StC{} and \CtS{} are forward and inverse Discrete Fourier Transformations;
in terms of the slot vector,
\StC{} maps $\coeffvec{m}$ to $\sigma(m)$, and \CtS{} maps $\sigma(m)$ to $\coeffvec{m}$.

For a power of two $n$, define the DFT matrix
\[
    \bm{V}_n=(\zeta_{4n}^{5^i j})_{i,j\in[n]}.
\]
Then $\bm{V}=\bm{V}_{N/2}$ satisfies $\sigma(m)=\bm{V}\coeffvec{m}$.
Thus, \StC{} evaluates $\bm{V}$ and \CtS{} evaluates $\bm{V}^{-1}$ in the slots.
For homomorphic evaluation, we use bit-reversed coefficient order:
\CtS{} outputs the coefficients in this order, and \StC{} accepts them in the same order.
The corresponding conversion matrices admit decompositions into
$\log(N/2)$ sparse layers~\cite{dft_decomp,dft_decomp2},
requiring $O(\log N)$ levels and ring operations.

\subsubsection{Homomorphic Sign Evaluation}
\label{sec:prelim:sign-cleaning}

We also introduce homomorphic sign evaluation~\cite{CKK20} for the cleaning-based bootstrapping.
The sign function can be approximated by a repeated composition of a fixed polynomial $f$.
Applying the approximation to $y/2$ gives the circuit $\operatorname{Sign}_q(y)$ below,
by specializing \cite[App.~D, Cor.~6]{CKK20}.
\begin{lemma}[Approximate Sign Evaluation]
    \label{lem:sign-cleaning}
    For every integer $q \ge 2$, there is a polynomial circuit $\operatorname{Sign}_q$
    of size and multiplicative depth $O(\log q)$ whose approximate evaluation satisfies
    \[
        \abs{\widehat{\operatorname{Sign}}_q(y) -\operatorname{sgn}(y)} \le 2^{-q}
    \]
    for every real $y$ with $1/2 \le \abs{y} \le 3/2$,
    provided each arithmetic gate introduces error of magnitude at most
    a sufficiently small constant multiple of $2^{-q}$.
\end{lemma}

\section{Encrypted Fixed-Point Multiplication by Focused Encoding}

We illustrate our fixed-point multiplication within the encrypted setting.
Its main ingredient is the focused encoding,
which homomorphically encodes a single real number with high precision into a polynomial with small coefficients.

\subsection{Focused Encoding}

We encode a single fixed-point number into the cyclotomic ring
$\calR = \bZ[X] / (X^N + 1)$ with power-of-two dimension $N$
using a `focal element' $u \in \calR$.

We control both the approximation error at the focus embedding and the magnitude at the remaining embeddings.
Hence, we say that $m \in \RealExt$ \emph{encodes} $z \in \bR$ within error $(\epsilon, B)$ at scale $\Delta > 0$ if
\[
  \abs{\tfrac{1}{\Delta} \sigma_0(m) - z} \le \epsilon, \qquad
  \abs{\tfrac{1}{\Delta} \sigma_j(m)} \le B \quad (j \neq 0).
\]

\subsubsection{Focal Element}

A \emph{focal element} is an element which is small at the focus embedding $\sigma_0$,
which gives more precision at the embedding with shaped error $u e$.
We say a focal element $u$ has $p$-bit \emph{gain} and $b$-bit \emph{overhead} if
\begin{equation}
  \label{eqn:gain-overhead}
  \log \abs{\sigma_0(u)} \le - p + b, \qquad
  \max_{j \neq 0} \log \abs{\sigma_j(u)} \le b,
\end{equation}
whose meaning is clarified in the following lemma.

\begin{lemma}[Error shaping by focal element]
  \label{lem:shaped-error}
  Suppose $u$ is given as in \eqref{eqn:gain-overhead}.
  Consider a plaintext $m \in \RealExt$ encoding $z \in \bR$ within error $(\epsilon, B)$,
  and a shaped error $u e$ with $2^b \canorm{e} \le \gamma \Delta$ for some $\gamma > 0$.
  Then, the plaintext $m + u e$ encodes $z$ within error $(\epsilon + 2^{-p} \gamma, B + \gamma)$.

  In particular, such an operation increases the coefficient size by at most $\gamma \Delta = \Omega(2^b)$,
  while having precision of roughly $p$-bit.
\end{lemma}
\begin{proof}
  This follows from
  $\abs{\sigma_j(u e)} \le \gamma \Delta$ for $j \neq 0$ and $\abs{\sigma_0(u e)} \le 2^{-p} \gamma \Delta$.
\end{proof}

We call a focal element $u$ is \emph{asymptotic} if its gain and overhead satisfies
\[
  p = \Omega(N \log N), \qquad b = O(\log N).
\]
As it grants superlinear gain in the ring dimension,
it will be a core ingredient of the focused encoding with large precision.

\begin{theorem}[Existence of a Focal Element]
    \label{thm:focal-existence}
    Fix $\beta>0$. For every sufficiently large power-of-two ring dimension $N$,
    there is an asymptotic focal element $u \in \calR \setminus \{0\}$
    which is conjugate-invariant, i.e., $u(X^{-1}) = u(X)$,
    and satisfies
    \[
        0 < u_F = \sigma_0(u) \le N^{-\beta N/2}, \qquad
        \canorm{u} = O\!\left(N^{\beta+1/2}\sqrt{\log N}\right).
    \]
\end{theorem}
\begin{proof}
    Let $c_0 = 1$ and $c_j = X^j + X^{-j}$ for $1 \le j < N/2$ be a conjugate-invariant basis.
    These satisfy $\abs{\sigma_k(c_j)} \le 2$.
    Set $B=\lceil N^\beta\rceil$ and $T=B\sqrt{4N\log_e(2N)}$,
    and for $\epsilon\in([-B,B]\cap\bZ)^{N/2}$, let $v_\epsilon = \sum_j \epsilon_j c_j$.
    We will show that there exist $\epsilon\ne\epsilon'$ such that
    \[
        \max\{\canorm{v_\epsilon},\canorm{v_{\epsilon'}}\}\le T,
        \qquad
        \abs{\sigma_0(v_\epsilon-v_{\epsilon'})}
            \le \frac{4T}{(2B+1)^{N/2}-2}.
    \]

    Choose $\epsilon$ uniformly at random.
    For each slot $k$, $\sigma_k(v_\epsilon) = \sum_j \epsilon_j \sigma_k(c_j)$
    is a sum of $N/2$ independent centered terms bounded by $2B$,
    so Hoeffding's inequality gives
    \[
        \Pr\left[\abs{\sigma_k(v_\epsilon)} > T\right]
        \le 2 \exp\left(-\frac{T^2}{2B^2 \sum_j \sigma_k(c_j)^2}\right)
        \le 2 \exp\left(-\frac{T^2}{4NB^2}\right) = \frac{1}{N}.
    \]
    A union bound over the $N/2$ slots gives $\Pr[\canorm{v_\epsilon} > T] \le 1/2$,
    so at least $(2B+1)^{N/2}/2$ distinct candidates satisfy $\canorm{v_\epsilon} \le T$.

    Their values at the $0$-th slot lie in $[-T,T]$, so two distinct candidates have difference
    $u=v_\epsilon-v_{\epsilon'}\in\calR\setminus\{0\}$ satisfying
    \[
        \abs{\sigma_0(u)} \le \frac{2T}{(2B+1)^{N/2}/2 - 1}, \quad
        \canorm{u} \le 2T = O\!\left(N^{\beta+1/2}\sqrt{\log N}\right).
    \]
    Substituting $B$ and $T$ gives
    \[
        \frac{4T}{(2B+1)^{N/2}-2}
        = O\!\left(2^{-N/2}N^{\beta+1/2}\sqrt{\log N}\right)N^{-\beta N/2}
        = o\!\left(N^{-\beta N/2}\right),
    \]
    proving the bound at the focus embedding for sufficiently large $N$.
    We may flip sign to obtain $\sigma_0(u)>0$.
\end{proof}

Finding such an element is a substantial task,
potentially taking $2^{O(N \log N)}$.
Hence, we admit exponential preprocessing for our algorithm.
Finding an asymptotic focal element efficiently is an interesting open problem.

\subsubsection{Focused Encoding}

Using a focal element $u$,
we encode a single real number at the focus embedding,
while shaping the rounding error by the element.
\begin{definition}[Focused Encoding]
  Given a real number $z \in \bR$ and scale factor $\Delta > 0$,
  the \emph{focused encoding} is defined as
  \[
    \Ecd^{[u]}(z; \Delta) = u \cdot \round{\Delta u^{-1} \cdot \sigma^{-1} (z \cdot \bm{e}_0)} \in \calR,
    \qquad
    \Dcd^{[u]}(m; \Delta) = \tfrac{1}{\Delta} \operatorname{Re} \sigma_0(m) \in \bR.
  \]
\end{definition}

\begin{lemma}[Encoding]
  Suppose $u$ is a focal element with $p$-bit gain and $b$-bit overhead,
  and $\gamma \Delta \ge 2^b \cdot E_\Ecd$ for $E_\Ecd = \tfrac{N}{2}$.
  Given $z \in \bR$,
  $\Ecd^{[u]}(z)$ encodes $z$ within error $(2^{-p} \gamma, \gamma)$.
\end{lemma}
\begin{proof}
  Let $m_0 = \Delta \sigma^{-1}(z \bm{e}_0) \in \RealExt$,
  which encodes $z$ within error $(0, 0)$.
  One has
  \[
    m := \Ecd^{[u]}(z; \Delta) = u \cdot (u^{-1} m_0 + e) = m_0 + u e
  \]
  where $\norm{e}_\infty \le \tfrac{1}{2}$, so that $\canorm{e} \le \tfrac{N}{2}$.
  The error bound follows from \Cref{lem:shaped-error}.
\end{proof}

\subsection{Efficient Encoding Algorithms}
\label{sec:construction:encoding}

A direct implementation of $\Ecd^{[u]}$ and $\Dcd^{[u]}$ use $n$-bit arithmetic on all $N$ coefficients,
incurring a nearly quadratic cost in $n$.
We give efficient algorithms of both maps using a tower of cyclotomic rings.
Encoding ascends the tower, increasing the ring dimension while reducing the coefficient precision.
Decoding descends the same tower, decreasing the ring dimension while increasing the coefficient bit lengths.
In both directions, the intermediate polynomials have total bit length $O(n)$,
yielding quasilinear complexity.

In this section, we will choose $u$ as in \Cref{thm:focal-existence} with $\beta=1$
for sufficiently large precision $n$,
taking the power-of-two dimension $N$ to be a sufficiently large
constant multiple of $n/\log n$.
Then $u$ has at least $n$-bit gain and $b=O(\log N)$-bit overhead.

\paragraph{Reformulating Encoding and Decoding.}
Consider $\Delta = 1$ for simplicity.
For encoding, notice that the core is $u \cdot \round{u^{-1} m}$ where $m \in \calR$ holds the scalar $z \in \bR$ at the focus embedding.
Instead, we consider computing the shaped rounding
\[
  u \cdot \round{u^{-1} z} = z - u \cdot [u^{-1} z]_1,
\]
then adjust the algorithm to output the desired encoding approximately.

For decoding, notice
\[
  \operatorname{Re}\sigma_0(m) = \tfrac{1}{2} \operatorname{Tr}_{\calK/\bQ} (m \cdot \sigma^{-1}(\bm{e}_0)).
\]
In particular, $\sigma_0(u)/u$ approximates $\sigma^{-1}(\bm{e}_0)$ with high precision,
so we can compute the following instead:
\begin{equation}\label{eq:trace-decoding}
  \frac{\sigma_0(u)}{2} \cdot \operatorname{Tr}_{\calK/\bQ}(u^{-1} m).
\end{equation}

Each of these computations can be performed while ascending and descending tower,
which admits intermediate representations of bit length $O(n)$.

\paragraph{Tower of Rings.}
For $0 \le i \le \log N$, define tower of rings
\[
    \calR_i=\bZ[X_i]/(X_i^{2^i}+1), \qquad
    \calK_i=\bQ[X_i]/(X_i^{2^i}+1),
\]
where $\calR_0 = \bZ$ and $\calR_{\log N} = \calR$,
and $\calR_{i-1}$ is a subring of $\calR_i$ by identification $X_{i-1} = X_i^2$.

Then, let $u_i = \operatorname{N}_{\calK/\calK_i}(u) \in \calR_i$ be a relative norm of $u$.
We have
\[
    u_{i-1}(X_{i-1}) = u_i(X_i) \cdot u_i(-X_i),
    \qquad 1\le i\le\log N,
\]
so roughly, $u^{-1} m = u_{\log N}^{-1} m$ can be computed iteratively from $u_0^{-1} m$ as
\begin{equation*}
  u_i^{-1}(X_i) \cdot m = u_i(-X_i) \cdot (u_{i-1}^{-1}(X_{i-1}) \cdot m).
\end{equation*}

\subsubsection{Fractional Ascent}
To compute $[u^{-1} z]_1$,
we ascend the tower while approximating $[u_i^{-1} z]_1$ by $t_i \in \bR[X_i]/(X_i^{2^i} + 1)$.
The details are given in \Cref{alg:fractional-ascent}.

\begin{algorithm}[htbp]
    \caption{Fractional Ascent}
    \label{alg:fractional-ascent}
    \begin{algorithmic}[1]
        \Require Precision $(p_i)_{i=0}^{\log N}$, scalar $z \in \bR$ and relative norms $(u_i)_{i=0}^{\log N}$.
        \Ensure An approximation to $[u^{-1} z]_1$, with coefficients in $[-1/2,1/2]$.
        \State $t_0\gets\left[z/u_0\right]_1$ \Comment{with $p_0$-bit precision}
        \For{$i=1,\ldots,\log N$}
            \State $t_i(X_i)\gets\left[u_i(-X_i)t_{i-1}(X_i^2)\right]_1$ \Comment{with $p_i$-bit precision}
        \EndFor
        \State \Return $t_{\log N}$
    \end{algorithmic}
\end{algorithm}

\begin{lemma}[Correctness of fractional ascent]
    \label{lem:fractional-ascent}
    For an exactly represented number $z\in\bR$, set precision as
    \[
        \ell=\left\lceil\log\norm{u}_1\right\rceil, \qquad
        p_i=\frac{N}{2^i}(\ell+1)+2.
    \]
    Then \Cref{alg:fractional-ascent} on input $z$ returns $t$ satisfying
    \[
        t=[u^{-1}z+e]_1, \qquad
        \norm{e}_\infty\le\frac{1}{8\norm{u}_1}.
    \]
\end{lemma}
\begin{proof}
    The relative-norm recursion gives $\norm{u_i}_1 \le \norm{u}_1^{N/2^i}\le2^{(N/2^i)\ell}$.
    We show inductively that
    \[
        t_i=[u_i^{-1}z+e_i]_1, \qquad
        \norm{e_i}_\infty\le2^{-p_i}.
    \]
    For $i=0$, this follows from the rounding error bound
    $2^{-p_0-1}\le2^{-p_0}$.
    For $i\ge1$, multiplication by the integral polynomial $u_i(-X_i)$
    preserves congruence modulo $1$.
    Hence the relative-norm identity gives $t_i=[u_i^{-1}z+e_i]_1$,
    where $e_i$ is $u_i(-X_i)e_{i-1}(X_i^2)$ plus the rounding error at level $i$.
    It yields
    \[
        \norm{e_i}_\infty
        \le \norm{u_i}_1\norm{e_{i-1}}_\infty + 2^{-p_i-1}
        \le2^{-p_i}.
    \]
    Taking $i=\log N$ gives $t=[u^{-1}z+e]_1$ with
    $\norm{e}_\infty\le2^{-\ell-3}\le1/(8\norm{u}_1)$.
\end{proof}

Shaped rounding follows directly from \Cref{lem:fractional-ascent}.
Writing $t=u^{-1}z+k+e$ with $k\in\calR$, we have
$z-ut=-uk-ue$ and $\norm{ue}_\infty\le1/8$.
Hence integer rounding gives
\[
    P=\round{z-ut}=-uk=z+u(e-t)\in u\calR,
\]
where $\norm{e-t}_\infty\le1/8+1/2=5/8$.

\paragraph{Complexity of Fractional Ascent.}
With the precisions specified in \Cref{lem:fractional-ascent},
$\ell=\lceil\log\norm{u}_1\rceil=O(\log N)$ for focused encoding, so level $i$ uses
$2^i$ coefficients represented with $O((N/2^i)\log N)$ bits each,
with total at $O(N \log N)$ bits.
By Kronecker substitution, each of its constant number of polynomial products
costs $O(\ModMultCost(N\log N))=O(\ModMultCost(n))$.
Taking fractional parts and rounding fit within the same bound.
There are $O(\log N)$ levels, giving total cost
\[
    O(\ModMultCost(n)\log N),
\]
assuming the relative norms are precomputed.
The initial scalar division and the final product by $u$
each cost $O(\ModMultCost(n))$.

\subsubsection{Focused Encoding by Fractional Ascent}

To keep the background embeddings bounded for repeated multiplication,
we multiply the shaped-rounding output by the precomputed polynomial
\[
    v=\round{4N\sigma^{-1}(\bm e_0)},\qquad v_F=\sigma_0(v),
\]
where rounding preserves conjugate invariance.

\begin{lemma}[Efficient focused encoding]
    \label{lem:efficient-encoding}
    Suppose $\log \Delta = O(\log N)$ and $\gamma\Delta\ge3\cdot2^b N^2$.
    For $\abs{z}\le1$, apply \Cref{alg:fractional-ascent}
    to $\Delta z/v_F$ with the precisions of \Cref{lem:fractional-ascent},
    and denote its output by $t$.
    Then
    \[
        \widetilde{\Ecd}^{[u]}(z;\Delta)=v\round{\Delta z/v_F-ut}
    \]
    encodes $z$ within error $(\gamma2^{-n},\tfrac{1}{7}+\gamma)$ at scale $\Delta$.
    Evaluation takes $O(\ModMultCost(n)\log n)$ time,
    with the relative norms, $v$, and $\Delta/v_F$ precomputed.
\end{lemma}
\begin{proof}
    Rounding gives
    \[
        v_F\ge\tfrac72N, \qquad
        \max_{j\ne0}\abs{\sigma_j(v)}\le\tfrac N2, \qquad
        \canorm{v}\le\tfrac92N.
    \]
    Scalar rounding gives
    \[
        \widetilde{\Ecd}^{[u]}(z;\Delta)=\Delta z\,v/v_F+u(vr),
        \qquad \norm{r}_\infty\le\tfrac58.
    \]
    The first term encodes $z$ within error $(0,1/7)$ at scale $\Delta$.
    Since $\canorm{vr}\le(9N/2)(5N/8)<3N^2$,
    \Cref{lem:shaped-error} gives the result.
    Scalar normalization and multiplication by $v$ add $O(\ModMultCost(n))$ work
    to the scalar-rounding cost above, giving total time
    $O(\ModMultCost(n)\log N)=O(\ModMultCost(n)\log n)$.
\end{proof}

\subsubsection{Trace Descent}

To evaluate \Cref{eq:trace-decoding}, we descend the tower using relative traces.
\Cref{alg:trace-descent} returns the integer
\[
    m_0=u_0\operatorname{Tr}_{\calK/\bQ}(u^{-1}m)
\]
exactly.

\begin{algorithm}[htbp]
    \caption{Trace Descent}
    \label{alg:trace-descent}
    \begin{algorithmic}[1]
        \Require Polynomial $m\in\calR$ and relative norms $(u_i)_{i=0}^{\log N}$.
        \Ensure The integer $u_0\operatorname{Tr}_{\calK/\bQ}(u^{-1}m)$.
        \State $m_{\log N}\gets m$
        \For{$i=\log N,\ldots,1$}
            \State $m_{i-1}\gets\operatorname{Tr}_{\calK_i/\calK_{i-1}}\bigl(m_i(X_i)u_i(-X_i)\bigr)$
        \EndFor
        \State \Return $m_0$
    \end{algorithmic}
\end{algorithm}

Correctness of \Cref{alg:trace-descent} follows from iterating the obvious identity
\[
    \frac{m_{i-1}}{u_{i-1}}
    =\operatorname{Tr}_{\calK_i/\calK_{i-1}}\left(\frac{m_i}{u_i}\right).
\]

\paragraph{Complexity of Trace Descent.}
Suppose $m$ has $O(\log N)$-bit coefficients.
Each relative trace increases the coefficient $\ell^1$-norm by at most a factor of two, so
\[
    \norm{m_i}_1
    \le\frac{N}{2^i}\norm{m}_1\prod_{j=i+1}^{\log N}\norm{u_j}_1
    \le\frac{N}{2^i}\norm{m}_1\norm{u}_1^{N/2^i-1}.
\]
Thus $m_i$ has $2^i$ coefficients of $O((N/2^i)\log N)$ bits each.
Each exact product costs $O(\ModMultCost(N\log N))=O(\ModMultCost(n))$,
giving $O(\ModMultCost(n)\log N)$ over the tower, with the relative norms precomputed.

\subsubsection{Focused Decoding by Trace Descent}
\label{sec:construction:decoding}

Using the trace descent, the decoder computes
\[
    \widetilde{\Dcd}^{[u]}(m;\Delta)
    =\frac{\sigma_0(u)}{2\Delta u_0}\,m_0
\]
to $n+1$ fractional bits using a precomputed approximation to the scalar factor,
and returns the result $\hat z$.
The integer $m_0$ has $O(N\log N)$ bits, so $n+O(N\log N)=O(n)$ fractional bits
of the scalar factor suffice to keep the final multiplication and rounding error within $2^{-n}$.
This step costs $O(\ModMultCost(n))$.

\begin{lemma}[Efficient focused decoding]
    \label{lem:efficient-decoding}
    Suppose $m\in\calR$ has $O(\log N)$-bit coefficients and encodes
    $z\in[-1,1]$ within error $(\epsilon,B)$ at scale $\Delta\ge1$.
    Then evaluating $\widetilde{\Dcd}^{[u]}(m;\Delta)$ returns $\hat z$ satisfying
    \[
        \abs{\hat z-z}\le\epsilon+(B+1)2^{-n}
    \]
    in $O(\ModMultCost(n)\log n)$ time,
    where the relative norms and scalar factor are precomputed.
\end{lemma}
\begin{proof}
    Since each $\sigma_j(u)$ is real, the two conjugate embeddings contribute
    $2\operatorname{Re}\sigma_j(m)/\sigma_j(u)$ to the trace. Hence,
    \[
        \widetilde{\Dcd}^{[u]}(m;\Delta)
        =\frac{\operatorname{Re}\sigma_0(m)}{\Delta}
        +\sum_{j\ne0}\frac{\sigma_0(u)}{\sigma_j(u)}
          \frac{\operatorname{Re}\sigma_j(m)}{\Delta}.
    \]

    Let $H=\canorm{u}$.
    Since $\operatorname{N}_{\calK/\bQ}(u)\ge1$, for each $j\ne0$,
    \[
        1\le\prod_{k=0}^{N/2-1}\abs{\sigma_k(u)}
        \le\abs{\sigma_0(u)\sigma_j(u)}H^{N/2-2}.
    \]
    Substituting bounds of \Cref{thm:focal-existence} implies
    \[
        \sum_{j\ne0}\left|\frac{\sigma_0(u)}{\sigma_j(u)}\right|
        \le N\abs{\sigma_0(u)}^2H^{N/2-2}
        \le 2^{-\Omega(N\log N)}.
    \]
    Taking the constant in the choice $N=\Theta(n/\log n)$
    sufficiently large makes this bound at most $2^{-n}$.
    The encoding bounds on $m$ therefore give
    $\abs{\widetilde{\Dcd}^{[u]}(m;\Delta)-z}\le\epsilon+B2^{-n}$.
    The final scalar multiplication and rounding add error at most $2^{-n}$,
    thus $\abs{\hat z-z}\le\epsilon+(B+1)2^{-n}$.

    The trace-descent and scaling costs above give total time
    $O(\ModMultCost(n)\log N)=O(\ModMultCost(n)\log n)$.
\end{proof}

\subsection{Levelled Multiplication by Focused Encoding}

Now, we present a focused CKKS scheme as a levelled homomorphic encryption scheme,
which preserves the shaped error.
As a result, we will obtain a levelled homomorphic multiplication.

\begin{theorem}[Levelled homomorphic fixed-point multiplication]
  \label{thm:mult-levelled}
  There is a multiplication algorithm
  \[
    \Mult^{[u]} : \calR_{Q_\ell}^2 \times \calR_{Q_\ell}^2
    \longrightarrow \calR_{Q_{\ell-1}}^2,
  \]
  with a suitable modulus chain $\{ Q_\ell \}_{\ell \in [L+1]}$ and a constant $0 < \gamma \le 1/4$ such that,
  whenever $\Dec(\ct_i)$ encodes $z_i \in [-1, 1]$ within error $(\epsilon_i, B_i)$ with $\epsilon_i, B_i \le \tfrac{1}{4}$,
  $\Dec(\Mult^{[u]}(\ct_1, \ct_2))$ encodes $z_1 z_2$ within error $(\epsilon_\times, B_\times)$
  where
  \[
    \epsilon_\times = \abs{z_1} \epsilon_2 + \abs{z_2} \epsilon_1 + \epsilon_1 \epsilon_2 + \gamma 2^{-n}, \qquad B_\times = B_1 B_2 + \gamma.
  \]
  Assuming the security parameter satisfies $\lambda L = O(n / \log^2 n)$,
  it runs in time $O(M(n L))$.
\end{theorem}

\subsubsection{Focused Encryption}

Fix modulus $Q$.
In RLWE with ternary secrets,
the secret key $\sk$ is sampled from the ternary distribution of hamming weight $h$,
and an RLWE sample is given as
$(b, a) \in \calR_Q^2$ where
$b = - a \cdot \sk + e$, $a$, $e$ are each sampled from uniform and discrete Gaussian distribution.
The decisional RLWE problem states that
the sample is indistinguishable from uniform,
providing secrecy for encryption schemes like CKKS.

If the focal element $u$ is invertible in $\calR_Q$,
multiplication by $u$ becomes a bijection,
so multiplying $u$ to the RLWE sample stays indistinguishable from uniform.
Hence, we define \emph{focused encryption} as
\[
  \Enc^{[u]}_\sk(m) = ([m]_Q, 0) + u \cdot (- a \cdot \sk + e, a), \qquad
  \Dec^{[u]}_\sk(\ct = (b, a)) = \Dec_\sk(\ct) = [b + a \cdot \sk]_Q,
\]
so that the encryption error follows the shape of $u$.
Specifically, one has
\[
  \Dec_\sk(\Enc^{[u]}(m)) = m + u e_\Enc \pmod{Q}, \qquad \canorm{e_\Enc} \le E_\Enc.
\]
By \Cref{lem:shaped-error}, it adds $(2^{-p} \gamma, \gamma)$ to the error bounds where $\gamma = 2^b E_\Enc / \Delta$.

On the other hand, a ciphertext $\ct$ satisfying $\Dec_\sk(\ct) = m \pmod{Q}$
decrypts to $m$ whenever $\norm{m}_\infty < \tfrac{Q}{2}$.
In particular, if $m \in \calR$ encoding $z \in [-1, 1]$ within error $(\epsilon, B)$,
one has
\[
  \norm{m}_\infty \le \canorm{m} \le \max(1 + \epsilon, B) \cdot \Delta,
\]
so $\max(1 + \epsilon, B) \cdot \Delta < Q/2$ suffices for decryption to recover the plaintext $m$.

Note that $u$ is invertible modulo $Q$ if and only if $Q$ is coprime with $\operatorname{N}_{\calK/\bQ}(u)$;
we call such a modulus $Q$ \emph{admissible}, and we only work with admissible moduli.
For an asymptotic focal element, $\log\abs{\operatorname{N}_{\calK/\bQ}(u)}=O(N\log N)$,
so only $O(N\log N)$ primes are not admissible.
Since prime number theorem guarantees $\Theta(T/\log T)$ primes in $[T,2T]$ for each $T$,
we can choose admissible primes large enough to satisfy our parameter requirements,
while keeping their bit lengths $O(\log N)$.

\paragraph{Security parameters.}
\label{par:security-parameters}
Let $\lambda$ be the security parameter,
$Q_{\max}$ be the largest modulus used in the construction,
including auxiliary key-switching moduli,
and $\sigma$ be the standard deviation of error distribution $\chi$.
In the small-error regime
$\log(Q_{\max}/\sigma)=\Theta(\log Q_{\max})$,
the dual-lattice and BKZ estimates discussed in~\cite{Albrecht17} give
\[
    \log\delta=\Theta\!\left(\frac{\log Q_{\max}}{N}\right),
    \qquad
    \log T_{\mathrm{BKZ}}=\Theta\!\left(
        \frac{N}{\log Q_{\max}}
        \log\frac{N}{\log Q_{\max}}
    \right).
\]
Thus choosing $N=\Omega(\lambda\log Q_{\max})$ makes this attack cost at least $2^{\Omega(\lambda)}$.
Meanwhile, the sparse-secret attack of~\cite[Section~5]{Albrecht17}
can reduce the dimension by a constant fraction at a
repetition cost of $2^{\Theta(h)}$ in the sparse regime.
We choose $h=\Omega(\lambda)$ so that this cost is $2^{\Omega(\lambda)}$.
Our parameter requirements are therefore
\[
    N=\Omega(\lambda\log Q_{\max}),\qquad h=\Omega(\lambda).
\]
Together, these conditions ensure an attack cost of
$2^{\Omega(\lambda)}$ under the stated lattice-attack heuristic.
As in CKKS~\cite{ckks17}, we additionally assume circular security
for the evaluation keys.
Note that the precomputed focal element depends only on $N$
and has an $O(N\log N)$-bit description;
we assume security against non-uniform adversaries
that receive it as advice.

\subsubsection{Homomorphic Operations}

Now, we give a focused version of two core operations of CKKS, key-switching and rescale,
whose error is shaped by $u$.

\begin{lemma}[Key-switching and Rescale]
  Suppose $q \mid Q$, $P \ge Q$.
  Then, the following algorithms
  \begin{align*}
    \KS^{[u]}_{\sk \to \sk'}(\ct) = u \cdot \KS_{\sk \to \sk'}([u^{-1}]_Q \cdot \ct), \qquad
    \RS^{[u]}_q(\ct) = u \cdot \RS_q([u^{-1}]_Q \cdot \ct)
  \end{align*}
  runs in $O(\PolyMultCost(N, \log (PQ)))$ time.
  Also, if $\Dec_\sk(\ct) = m \pmod{Q}$, they satisfy
  \begin{align*}
    \Dec_{\sk'}(\KS^{[u]}_{\sk \to \sk'} (\ct)) &= m + u e_\KS \pmod{Q}, \\
    \Dec_\sk(\RS^{[u]}_q (\ct)) &= \tfrac{m}{q} + u e_\RS \pmod{Q/q}.
  \end{align*}
  with $\canorm{e_\KS} \le E_\KS$ and $\canorm{e_\RS} \le E_\RS$.
\end{lemma}
\begin{proof}
  The correctness follows from applying
  $\Dec([u^{-1}]_Q \cdot \ct) = [u^{-1}]_Q \cdot \Dec_\sk(\ct) \pmod{Q}$
  into \Cref{lem:ckks-ks-rs}, then multiplying $u$.
\end{proof}

Then, the levelled multiplication can be constructed as follows.
\begin{proof}[Proof of \Cref{thm:mult-levelled}]
  Consider ring dimension $N$ with $n = \Theta(N \log N)$
  with an asymptotic focal element $u$ with gain $p = n$ and background $b = O(\log N)$.
  We take modulus chain $\{ Q_\ell \}_{\ell \in [L+1]}$ and scale factor $\Delta_0$ as
  \[
    Q_\ell = q_0 \Delta_0^\ell, \qquad
    \Delta_0 \ge (2^b/\gamma) \cdot (E_\KS + E_\RS), \qquad
    q_0 \ge 4 \Delta_0.
  \]
  Note that both $\log \Delta_0$ and $\log q_0$ can be taken as $O(\log N)$.
  Taking auxiliary key-switching modulus comparable to the largest ciphertext modulus
  gives $\log Q_{\max}=O(L\log N)$.
  Since $n = \Theta(N\log N)$, the security assumption ensures $N = \Omega(\lambda \log Q_{\max})$.

  Given ciphertexts $\ct_i = (b_i, a_i) \in \calR_{Q_\ell}^2$,
  the multiplication algorithm is given as
  \begin{align*}
    \Mult^{[u]}(\ct_1, \ct_2)
    &= \RS^{[u]}_\Delta \Bigl((d_0, d_1) + \KS^{[u]}_{\sk^2 \to \sk}((0, d_2)) \Bigr) \in \calR_{Q_{\ell-1}}, \\
    (d_0, d_1, d_2) &= (b_1 b_2, a_1 b_2 + a_2 b_1, a_1 a_2),
  \end{align*}
  which runs in $O(\PolyMultCost(N, \log(P Q_\ell))) = O(\ModMultCost(N L \log N)) = O(\ModMultCost(n L))$ time.

  Now, we track its error. For $m_i = \Dec(\ct_i)$, we obtain
  \[
    \Dec(\Mult^{[u]}(\ct_1, \ct_2)) = \frac{1}{\Delta_0} m_1 m_2 + u \Bigl( \frac{e_\KS}{\Delta_0} + e_\RS \Bigr) \pmod{Q_{\ell-1}}.
  \]
  $m_1 m_2$ encodes $z_1 z_2$ within error
  $(\abs{z_1} \epsilon_2 + \abs{z_2} \epsilon_1 + \epsilon_1 \epsilon_2, B_1 B_2)$,
  and the key-switching and rescale error adds $(2^{-p} \gamma, \gamma)$ by \Cref{lem:shaped-error}.
\end{proof}

\subsection{Bootstrapping for Focused Encoding}

The levelled multiplication eventually reaches the base modulus, where it cannot continue.
Hence, we present the bootstrapping for Focused CKKS, which recovers the lost modulus,
supporting unbounded number of multiplications.
\begin{theorem}[Bootstrapping for Focused CKKS]
  \label{thm:bootstrapping-focused}
  Fix $L = \Theta(1)$ and a constant $0<\gamma\le1$.
  Assume $\lambda\ge2$ and $\lambda=O(n/\log^3 n)$ with a sufficiently small implied constant.
  With a suitable modulus chain $\{ Q_\ell \}_{\ell \in [L+1]}$,
  there is a bootstrapping algorithm
  \[
    \BTS^{[u]} : \calR_{Q_0}^2 \longrightarrow \calR_{Q_L}^2,
  \]
  such that whenever $\ct$ encrypts $z\in[-1,1]$ within error $(\epsilon, B)$
  with $\epsilon\le1$ and $B\le2$,
  its output encrypts $z$ within error $(\epsilon + \gamma 2^{-n}, B + \gamma)$.
  Substituting $\ModMultCost(b)=O(b\log b)$, the running time is
  \[
    O\!\left(n\log n+n(\log\lambda)^3\log\log\lambda\right).
  \]
\end{theorem}

Taking $L = 1$ and performing bootstrapping after each multiplication,
with suitable choice of $\gamma$ for each step, gives the following.
\begin{corollary}[Fixed-point multiplication for ciphertexts]
  \label{thm:mult-fhe}
  There is a multiplication algorithm
  \[
    \Mult^{[u]} : \calR_Q^2 \times \calR_Q^2
    \longrightarrow \calR_Q^2,
  \]
  with a suitable modulus $Q$ and a constant $0 < \gamma \le \tfrac{1}{4}$ such that,
  whenever $\ct_i$ encrypts $z_i \in [-1, 1]$ within error $(\epsilon_i, B_i)$ with $\epsilon_i, B_i \le \tfrac{1}{4}$,
  its output encrypts $z_1 z_2$ within error $(\epsilon_\times, B_\times)$ where
  \[
    \epsilon_\times = \abs{z_1} \epsilon_2 + \abs{z_2} \epsilon_1 + \epsilon_1 \epsilon_2 + \gamma 2^{-n}, \qquad B_\times = B_1 B_2 + \gamma.
  \]
  Using $\ModMultCost(b)=O(b\log b)$ and assuming $\lambda\ge2$ and $\lambda=O(n/\log^3 n)$, it runs in time
  \[
    O\!\left(n\log n+n(\log\lambda)^3\log\log\lambda\right).
  \]
\end{corollary}
\begin{proof}
  Correctness follows from levelled multiplication and
  \Cref{thm:bootstrapping-focused}, allocating a constant fraction of
  $\gamma$ to each operation.
  The multiplication adds $O(\ModMultCost(n))$ time, so the total running
  time has the same bound as bootstrapping.
\end{proof}

\subsubsection{Shaped Bootstrapping}

To obtain the desired bootstrapping, the resulting error should be extremely small in the focus slot,
so usual bootstrapping error does not suffice;
regarding usual error as shaped error will blow up at the focus embedding, since $u^{-1}$ has size $2^{\Omega(n)}$ at the embedding.

Instead, observe that if a ciphertext $\ct \in \calR_{\rho Q}^2$ is encrypting
\[
  \Dec(\ct) = \rho m + e, \qquad m \in \calR,
\]
multiplying $u [u^{-1}]_Q$ does not change the message $m$, while it turns the error into a shaped error $u \cdot ([u^{-1}]_Q e)$.
Crucially, $[u^{-1}]_Q$ has size upper bound of $Q$.

To integrate this with the bootstrapping circuit,
we consider a \emph{scaled bootstrapping} $\BTS_\rho$,
which outputs a ciphertext at modulus $\rho Q$ encrypting $\rho m+e_\BTS$
with $\canorm{e_\BTS}/\rho\le1/Q$.
Then, we define a \emph{shaped bootstrapping} as
\[
  \BTS^{[u]}(\ct) = u \cdot \RS_\rho([u^{-1}]_Q \cdot \BTS_\rho(\ct)),
\]
which scales the message back down while shaping the error.

In particular,
\[
  \Dec(\BTS^{[u]}(\ct)) = m + u e, \qquad
  e = [u^{-1}]_{Q} \cdot \tfrac{e_\BTS}{\rho} + e_\RS
\]
where $e_\BTS$ is the scaled bootstrapping error and $e_\RS$ is the rescale error.
Hence, sufficiently large $\rho$ can give a reasonable bound on the resulting error $e$.

\subsubsection{Scaled Bootstrapping}
\label{sec:scaled-bts}

We use a CKKS bootstrapping implementation with scaling added in \ModRaise{}.

\begin{proposition}[Scaled bootstrapping]
    \label{prop:scaled-bts}
    Fix a constant $L\ge1$ and a sufficiently large ring dimension $N$.
    Let moduli
    $Q_L=Q_0\Delta_0^L$,
    where $\Delta_0,Q_0$ are positive integers with
    $\Delta_0\ge N$ and $Q_0\ge2\Delta_0^2$.
    Suppose the secret is ternary of weight $h=\Theta(\lambda)\le N$,
    and the RLWE error is bounded by $\Sigma=Q_0^{O(1)}$, and
    $N/\log N=\Omega(\lambda\log Q_0)$.

    For a suitable integer $\rho=Q_L(N\Sigma)^{O(1)}$,
    there is a scaled bootstrapping algorithm
    that maps $\ct\in\calR_{Q_0}^2$
    encrypting $m=\Dec_\sk(\ct)$ with $\norm{m}_\infty\le2\Delta_0$,
    to $\ct'\in\calR_{\rho Q_L}^2$ satisfying
    \[
        \Dec_\sk(\ct')=\rho m+e_\BTS\pmod{\rho Q_L},
        \qquad \canorm{e_\BTS}\le\rho/Q_L.
    \]
    It runs in
    \[
        O\!\left((\log N+\log\lambda\log\log Q_0)
        \ModMultCost(N\log N\log Q_0)\right)
    \]
    time.
\end{proposition}

We defer the proof of \Cref{prop:scaled-bts}
to the end of this subsection.

\subsubsection{Subring Bootstrapping}

While the bootstrapping construction above would give quasilinear performance,
it has many logarithmic factors, which is not ideal for our purpose.
Instead, we split the large ciphertext into several small ciphertext over a subring,
reducing both the number and cost of ring operations needed for bootstrapping.

Consider a subring $\calR_0 = \bZ[Y]/(Y^{N_0} + 1)$ of $\calR$ by identification $Y = X^{N/N_0}$.
As observed in~\cite{BCKPS23},
\[
  \calR = \bigoplus_{j \in [N/N_0]} \calR_0 X^j
\]
as $\calR_0$-module.
Hence, when the secret key lies in $\calR_0$,
a ciphertext $\ct = (b, a)$ over $\calR$ splits into $N/N_0$ ciphertexts $\ct_j = (b_j, a_j)$ over $\calR_0$,
and its security is determined by the subring degree $N_0$.

Since we are bootstrapping the coefficients of the message polynomial,
we may bootstrap each subring ciphertext instead,
reducing the task to bootstrapping $\tfrac{N}{N_0}$ ciphertexts of ring dimension $N_0$.

We now prove \Cref{thm:bootstrapping-focused} using bootstrapping over this subring.
\begin{proof}[Proof of \Cref{thm:bootstrapping-focused}]
  The bootstrapping algorithm consists of the following steps.
  \begin{enumerate}
    \item Split $\ct$ into ciphertexts $\{\ct_j\}_{j\in[N/N_0]}$ over $\calR_0$.
    \item Bootstrap each component as $\ct'_j=\BTS_\rho(\ct_j)$ over $\calR_0$.
    \item Recombine the outputs as $\ct'=\sum_{j\in[N/N_0]}X^j\ct'_j$ over $\calR$.
    \item Apply the shaped bootstrapping transformation to get
      $\ct''=u\cdot\RS_\rho([u^{-1}]_{Q_L}\cdot\ct')$.
  \end{enumerate}

  We choose the secret key in $\calR_0$ with Hamming weight $h=\Theta(\lambda)$,
  and use it for both multiplication and bootstrapping.
  Choose ring dimension $N_0$ as power-of-two with
  \[
    N_0=\Theta(\lambda\log Q_0\log(\lambda\log Q_0))
  \]
  to satisfy the security bound $N_0/\log N_0=\Omega(\lambda\log Q_0)$.
  The bound $\lambda=O(n/\log^3 n)$ ensures $N_0\le N$.

  For the moduli chain, choose admissible $\Delta_0,Q_0$ with
  $\log\Delta_0,\log Q_0=\Theta(\log N)$ such that
  \[
    \gamma\Delta_0\ge 2^b(E_\KS+E_\RS),\qquad
    Q_0\ge2\Delta_0^2.
  \]
  and set $Q_\ell=Q_0\Delta_0^\ell$.
  Here, $E_\KS$ and $E_\RS$ are the error bounds for the full ring $\calR$.

  Now, let $m=\Dec_\sk(\ct)$. By the input bounds, one has
  \[
    \norm{m}_\infty\le\canorm{m}
    \le\max(1+\epsilon,B)\Delta_0\le2\Delta_0.
  \]
  After splitting, each component has the same coefficient bound,
  so they satisfy the input condition of \Cref{prop:scaled-bts}.
  Our scale choice also yields $\Sigma = O(\Delta_0) = Q_0^{O(1)}$.

  Applying the proposition for ring dimension $N_0$
  to each component with the scaling factor
  $\rho=Q_LQ_0^{O(1)}$ gives bootstrapping error at most $\rho/Q_L$ per component.
  Each component has the same canonical norm when viewed in $\calR$,
  and $\canorm{X^j}=1$, so recombination gives
  \[
    \Dec_\sk(\ct')=\rho m+e'\pmod{\rho Q_L},
    \qquad \frac{\canorm{e'}}{\rho}\le\frac{N}{N_0Q_L}.
  \]

  Then, after the shaped bootstrapping transformation,
  \[
    \Dec_\sk(\ct'')=m+ue''\pmod{Q_L},
    \qquad e''=[u^{-1}]_{Q_L}\frac{e'}{\rho}+e_\RS.
  \]
  Using $\canorm{[u^{-1}]_{Q_L}}\le NQ_L/2$, we obtain
  \[
    \canorm{e''}\le\frac{N^2}{2N_0}+E_\RS\le E_\KS.
  \]
  Our scale choice and \Cref{lem:shaped-error} therefore give
  output error $(\epsilon+\gamma2^{-n},B+\gamma)$.

  As for cost, bootstrapping all $N/N_0$ components costs
  \[
    O\!\left(\frac{N}{N_0}
      (\log N_0+\log\lambda\,\log\log Q_0)
      \ModMultCost(N_0\log N_0\log Q_0)\right).
  \]
  Since $\log Q_0=\Theta(\log N)$ and $n=\Theta(N\log N)$,
  we have $N_0=\Theta(\lambda\log n\log(\lambda\log n))$ and
  $\log N_0=\Theta(\log(\lambda\log n))$.
  Substituting, the cost is
  \[
    O\!\left(
    \frac{n\log\lambda\,\log\log n}
         {\lambda\log^2 n\,\log(\lambda\log n)}
    \ModMultCost(\lambda\log^2 n\,\log^2(\lambda\log n))\right).
  \]

  Splitting, recombination, and error shaping use moduli at most $\rho Q_L = N^{O(1)}$,
  adding $O(\ModMultCost(N\log N))=O(\ModMultCost(n))$ time.
  Substituting $\ModMultCost(b)=O(b\log b)$,
  the total cost is
  \[
    O\!\left(n\log n + n \log\lambda \log\log n \, (\log \lambda + \log\log n)^2\right).
  \]
  If $\log \lambda \le (\log n)^{1/4}$, the second term is $O(n\log n)$.
  Otherwise, $\log\log n < 4 \log \log \lambda$ and the second term is $O(n (\log \lambda)^3 \log \log \lambda)$,
  proving the claimed bound.
\end{proof}

\subsubsection{Proof of Scaled Bootstrapping}

For the scaled boostrapping, we follow the cleaning-based approach of~\cite{CKSS25,CL26},
which computes and removes the overflow introduced by modular raising using approximate sign evaluation.

\paragraph{Approximate modular reduction.}
We first give a polynomial circuit that extracts each bit of the overflow to constant accuracy,
refines them independently, and subtracts the reconstructed overflow.

\begin{lemma}[Approximate modular reduction]
\label{lem:adjacent-bit-reduction}
Let \(K,p\ge 1\) be integers, and let \(\ell=\lceil\log(2K+1)\rceil\).
Given \(t=I+x\) where \(I\in[-K,K]\cap\bZ\) and \(x \in [-\tfrac{1}{4}, \tfrac{1}{4}]\),
a polynomial arithmetic circuit using complex arithmetic and conjugation
returns \(x\) to absolute error at most \(2^{-p}\).
Its size and multiplicative depth are
\[
  O\bigl(\ell\log(p+\ell)\bigr)
  \qquad\text{and}\qquad
  O\bigl(\ell+\log(p+\ell)\bigr),
\]
respectively.
These bounds also hold under approximate evaluation when
each arithmetic gate incurs error of magnitude at most
a sufficiently small constant multiple of \(2^{-(p+\ell)}\).
\end{lemma}

\begin{proof}
Write \(I+K=\sum_{k \in [\ell]} 2^k b_k\), where \(b_k\in\{0,1\}\),
and put \(s_k=(-1)^{b_k}\).
We will recover the bits from the signs as $b_k = (1 - s_k)/2$.

First, we consider the exact identities to approximate.
For each \(k \in [\ell]\), let
\[
 v_k=\exp(2\pi i\phi_k),\qquad
 \phi_k=\left\{\frac{t+K+1/2}{2^{k+1}}\right\},
\]
where \(\{ y \} = y - \lfloor y \rfloor \in [0, 1)\) denotes the fractional part.
Since $x + 1/2 \in (0, 1)$,
the first binary digits of $\phi_k$ are $b_k, b_{k-1}, \ldots, b_0$.

For $k \ge 1$, consider
\[
  w_k=\exp\!\left(2\pi i\left(\phi_k-\frac{b_{k-1}}4+\frac18\right)\right),
\]
whose argument lies in \([\pi/4, 3\pi/4]\) if $b_k = 0$,
and in \([5\pi/4, 7\pi/4]\) if $b_k = 1$.
Thus \(w_k\) is within angle \(\pi/4\) of \(s_k i\),
and
\(s_k\operatorname{Im} w_k \in [1/\sqrt{2}, 1]\).
For \(w_0=v_0\), the same condition holds for $k = 0$.

Since
\[
 \exp\!\left(2\pi i\left(\frac18-\frac{b_{k-1}}4\right)\right)
   =\frac{1+i s_{k-1}}{\sqrt2},
\]
the signs can be recovered successively as
\[
  w_0=v_0, \qquad
  w_k = \frac{1+i s_{k-1}}{\sqrt2} \cdot v_k \quad(1\le k<\ell), \qquad
  s_k=\operatorname{sgn}(\operatorname{Im}w_k).
\]

In the actual approximation,
we approximate $v_{\ell - 1}$ by a polynomial,
then obtain approximations to $v_{\ell-2}, \ldots, v_0$ by repeated squaring,
using $v_{k-1} = v_k^2$.
Let $\widetilde{v}_k$ be the resulting approximations.
Then, we use the sign circuit \(\operatorname{Sign}_q\) of \Cref{lem:sign-cleaning}.
For a constant precision $p_0$, we compute the signs as
\begin{equation*}
  \begin{aligned}
   \widetilde w_0 &\gets \widetilde v_0,\\
   \widetilde w_k &\gets
     \frac{1+i\,\widetilde s_{k-1}}{\sqrt2}\,\widetilde v_k
         &&(1\le k<\ell),\\
   \widetilde s_k &\gets \operatorname{Sign}_{p_0}(\operatorname{Im}\widetilde w_k)
         &&(0\le k<\ell).
  \end{aligned}
\end{equation*}
We then refine the signs independently through
\[
  \widehat s_k\gets\operatorname{Sign}_{p+\ell}(\widetilde s_k),
\]
and return the remainder as
\[
 \widehat x\gets t+K-\sum_{k=0}^{\ell-1}2^k\,\frac{1-\widehat s_k}{2}.
\]

Now, we show that the error bounds hold with a choice on the precision.
The degree-$O(\ell)$ Taylor approximation computes the exponential $v_{\ell-1}$
with error at most a sufficiently constant multiple of $2^{-\ell}$,
as given in \cite{EC:CHKKS18}.
Evaluating it with $O(\ell)$ bits of precision
and applying $v_{k-1} = v_k^2$ yields approximations $\widetilde{v}_k$ with
$\abs{\widetilde{v}_k - v_k} \le \delta$
for a sufficiently small constant $\delta > 0$.
This step uses $O(\ell)$ gates and depth.

Then, we show inductively that $\abs{\widetilde{s}_k - s_k} \le \delta$.
For $k = 0$, this follows from $s_0 \operatorname{Im} v_0 \ge 1/\sqrt{2}$ and the approximation bounds.
For $k \ge 1$, the induction hypothesis and approximation bounds yield
\[
  \abs{\widetilde{w}_k - w_k}
  \le \frac{\abs{\widetilde{s}_{k-1} - s_{k-1}}}{\sqrt{2}} + O(\delta) = O(\delta).
\]
Since $s_k \operatorname{Im} w_k \in [1/\sqrt{2}, 1]$,
the approximate input satisfies $s_k \operatorname{Im} \widetilde{w}_k \in [1/2, 3/2]$ for sufficiently small $\delta$.
Thus \Cref{lem:sign-cleaning} gives $\abs{\widetilde{s}_k - s_k} \le 2^{-p_0} \le \delta$,
completing the induction.
Here, extracting the $\ell$ signs sequentially at constant precision
costs $O(\ell)$ size and depth.

After that, refining each sign gives
\(\lvert\widehat s_k-s_k\rvert\le 2^{-(p+\ell)}\)
with \(O(\log(p+\ell))\) size and depth per sign.
Hence, the errors in the refined signs contributes at most
$\sum_{k \in [\ell]} 2^{k-1} \, \abs{\widehat{s}_k - s_k} < 2^{-(p + 1)}$
to the error in $\hat{x}$.
Computing the weighted sum introduces a further error bounded by a sufficiently small constant multiple of $2^{-p}$,
so the total error is at most $2^{-p}$.
Summing up the size and depth gives the stated complexity.
\end{proof}

\begin{proof}[Proof of \Cref{prop:scaled-bts}]
    After \ModRaise{}, the lifted ciphertext satisfies
    $b+a\cdot\sk=m+Q_0I$
    with $\norm{I}_\infty\le K=O(h)$.
    Multiply by $\rho$ and use working scale $S=\rho Q_0$.
    Applying \CtS{} and separating real and imaginary parts gives coordinates
    $I_j+m_j/Q_0$, where $\abs{m_j/Q_0} \le 1/\Delta_0 < 1/8$.
    Apply \Cref{lem:adjacent-bit-reduction}, recombine the real and imaginary
    parts, and apply \StC{}.
    At scale $S$, the resulting coefficients $m_j/Q_0$ represent the desired message $\rho m$.

    Now, we show the error bounds.
    The \CtS{} error is clearly bounded by $(N \Sigma)^{O(1)}$.
    For \Cref{lem:adjacent-bit-reduction}, take
    \[
        p = \log(NQ_0Q_L) + O(1), \qquad
        \ell=\lceil\log(2K+1)\rceil.
    \]
    Since $2^{p+\ell}=O(NKQ_0Q_L)$,
    choosing $\rho=Q_L(N\Sigma)^{O(1)}$ sufficiently large meets the error tolerance of the lemma,
    and also keeps the \CtS{} error small relative to $S$.
    Hence, it recovers each $m_j/Q_0$ with error $O(2^{-p})$.
    The \StC{} matrix has $\ell^\infty$ operator norm at most $N$,
    and its fresh evalaution error is small relative to $S$,
    so the output error satisfies
    \[
        \frac{\canorm{e_\BTS}}{\rho}
        =O(NQ_0 2^{-p})\le\frac{1}{Q_L}.
    \]

    Since $p,\log S=O(\log Q_0)$ and $\ell=O(\log\lambda)$,
    the DFTs in \CtS{}, \StC{} and the modular reduction circuit together use
    \[
        D=O(\log N+\log\log Q_0),\qquad
        T=O\!\left(\log N+\log\lambda\log\log Q_0\right)
    \]
    levels and ring operations. The security assumption implies
    $\log Q_0=O(N)$, so $D=O(\log N)$.
    Meanwhile, the maximum modulus including the key-switching auxiliary modulus should be
    \[
        \log Q_{\max}
        =O\!\left(D\log S+\log(\rho Q_L)\right)
        =O(\log N\log Q_0).
    \]
    Thus the assumed bound on $N$ ensures $N=\Omega(\lambda\log Q_{\max})$.
    Each ring operation costs $O(\ModMultCost(N\log N\log Q_0))$,
    proving the claimed running time.
\end{proof}

\section{Generalization to Multiple Focus Slots and Precisions}
\label{sec:construction}

We extend the single-slot construction to cover multiple slots with varying precisions.
Then, we give constructions of focal elements realizing the required profiles.
Lastly, we present the focused slot bootstrapping, which is efficient in the concrete regime.

\subsection{Focused CKKS}
\label{sec:construction:general}

We extend the single-slot construction by allocating a separate precision at each slot.
We say that $m\in\RealExt$ \emph{encodes} $\bm z\in\bC^{N/2}$
within error $(\epsilon_j)_{j\in[N/2]}$ at scale $\Delta>0$ if
\[
    \abs{\tfrac{1}{\Delta}\sigma_j(m)-z_j}\le\epsilon_j,
    \qquad j\in[N/2].
\]

We generalize focal elements as follows.
\begin{definition}[Focal element]
\label{def:general-focal}
Given gains $p_j\ge0$ for $j\in[N/2]$, we say that $u\in\calR\setminus\{0\}$
is a \emph{focal element} with \emph{$p_j$-bit gain at slot $j$}
and common \emph{$b$-bit overhead} if
\[
    \abs{\sigma_j(u)}\le2^{b-p_j},
    \qquad j\in[N/2].
\]
\end{definition}

\begin{lemma}[Error shaped by focal element]
    \label{lem:general-shape}
    Suppose $u$ is given as in \Cref{def:general-focal}.
    Consider a plaintext $m \in \RealExt$ which encodes $\bm z \in \bC^{N/2}$
    within error $(\epsilon_j)_{j\in[N/2]}$ at scale $\Delta$.
    Introducing a shaped error $u e$ with $2^b \canorm{e} \le \gamma \Delta$,
    the resulting plaintext $m+ue$ encodes $\bm z$ within error
    $(\epsilon_j+2^{-p_j}\gamma)_{j\in[N/2]}$ at the same scale.
\end{lemma}
\begin{proof}
    For each $j\in[N/2]$,
    \[
        \frac{\abs{\sigma_j(ue)}}{\Delta}
        \le\frac{2^{b-p_j}\canorm{e}}{\Delta}
        \le2^{-p_j}\gamma.
    \]
    Then, the bounds follow by the triangle inequality.
\end{proof}

Hence, the common overhead $b$ controls the scale needed to absorb shaped error,
while the gains $p_j$ determine the precision at each slot.
These parameters are constrained by the field norm:
\[
    0\le\tfrac12\log\operatorname{N}_{\calK/\bQ}(u)
    =\sum_j\log\abs{\sigma_j(u)}
    \le\tfrac N2 b-\sum_j p_j.
\]
Thus, the overhead is larger than the average of gains, i.e.,
\[
    b\ge\frac2N\sum_j p_j.
\]

\subsubsection{Scheme Construction}

With the generalized focal element,
most operations are inherited verbatim from the single-slot case.
We record the scheme as follows.
\begin{itemize}[leftmargin=*]
    \item $\mathsf{KeyGen}(1^\lambda)$. Same as CKKS.
    \item $\Ecd^{[u]}(\bm z; \Delta)$. Given $\bm{z} \in \bC^{N/2}$ and scaling factor $\Delta$,
        returns $u \cdot \round{\Delta u^{-1} \sigma^{-1}(\bm{z})}$.
    \item $\Dcd^{[u]}(\pt; \Delta)$. Given $\pt \in \calR$, returns $\sigma(\pt)/\Delta \in \bC^{N/2}$.
    \item $\Enc^{[u]}_\sk(\pt)$. Given $\pt \in \calR$,
        samples a uniform mask $a \in \calR_Q$ and an error $e_\Enc \gets \chi$,
        and returns $(\pt, 0) + u \cdot (- a \cdot \sk + e_\Enc, a)$.
    \item $\Dec^{[u]}_\sk(\ct)$. Given $\ct = (b, a) \in \calR_Q^2$, returns $[b + a \cdot \sk]_Q$.
    \item $\KS^{[u]}_{\sk\to\sk'}(\ct)$.
        Given $\ct \in \calR_Q^2$,
        returns $u \cdot \KS_{\sk\to\sk'}([u^{-1}]_Q \ct) \in \calR_Q^2$.
    \item $\RS_q^{[u]}(\ct)$. Given $\ct \in \calR_Q^2$ with $q\mid Q$, returns $u \cdot \RS_q([u^{-1}]_Q\ct)\in\calR_{Q/q}^2$.
    \item $\Add^{[u]}(\ct_1,\ct_2)$ returns $\ct_1+\ct_2 \in \calR_Q^2$.
    \item $\Mult^{[u]}(\ct_1, \ct_2)$.
        Given $\ct_i = (b_i, a_i) \in \calR_{Q}^2$,
        computes $(d_0, d_1, d_2) = (b_1 b_2, a_1 b_2 + a_2 b_1, a_1 a_2)$
        and returns
        \[
            \RS^{[u]}_\Delta((d_0, d_1) + \KS^{[u]}_{\sk^2 \to \sk}((0, d_2))) \in \calR_{Q/\Delta}^2.
        \]
    \item $\PtMult^{[u]}(\pt, \ct)$.
        Given $\ct \in \calR_Q^2$ and $\pt \in \calR$,
        returns $\RS^{[u]}_\Delta(\pt \cdot \ct)$.
    \item $\Aut^{[u]}_\kappa(\ct)$.
        Given $\ct = (b, a)$ and $\kappa \in \Gal(\calK/\bQ)$,
        returns $\KS^{[\kappa(u)]}_{\kappa(\sk) \to \sk}((\kappa(b), \kappa(a)))$.
\end{itemize}

\begin{lemma}[Encoding]
    \label{lem:general-encoding}
    Suppose $u$ is given as in \Cref{def:general-focal},
    and $\gamma \Delta \ge 2^b \cdot E_\Ecd$ for $E_\Ecd = \tfrac{N}{2}$.
    Given $\bm z \in \bC^{N/2}$,
    $\Ecd^{[u]}(\bm z;\Delta)$ encodes $\bm z$ within error
    $(2^{-p_j}\gamma)_{j\in[N/2]}$ at scale $\Delta$.
\end{lemma}
\begin{proof}
    Let $m_0=\Delta\sigma^{-1}(\bm z)$, which encodes $\bm z$
    exactly at scale $\Delta$.
    By the definition of the encoder,
    \[
        \Ecd^{[u]}(\bm{z};\Delta)
        =u\round{u^{-1}m_0}=m_0+ue,
        \qquad \norm{e}_\infty\le\tfrac12.
    \]
    Hence $\canorm{e}\le N/2=E_\Ecd$, and the claim follows
    from $2^b\canorm{e}\le\gamma\Delta$ and \Cref{lem:general-shape}.
\end{proof}

With multiple slots and complex messages, automorphisms are now given a meaning.
Automorphisms transform the shape of the error, as we illustrate below.

\begin{lemma}[Automorphism]
    \label{lem:automorphism}
    Consider an automorphism $\kappa \in \Gal(\calK/\bQ)$.
    For $\ct \in \calR_Q^2$, $\Aut^{[u]}_\kappa(\ct)$ introduces error shaped by $\kappa(u)$, i.e.,
    \[
        \Dec_\sk(\Aut^{[u]}_\kappa(\ct)) = \kappa(\Dec_\sk(\ct)) + \kappa(u) \cdot e_\KS \pmod{Q}.
    \]
\end{lemma}
\begin{proof}
    Writing $\ct=(b,a)$, applying $\kappa$ to both components gives
    \[
        \kappa(b)+\kappa(a)\kappa(\sk)
        =\kappa(b+a \cdot \sk)
        =\kappa(\Dec_\sk(\ct))\pmod Q.
    \]
    Applying \Cref{lem:ckks-ks-rs} gives the result.
\end{proof}

In particular, the error shape is preserved when $\kappa$ fixes $u$.
We will use subring elements to obtain rotations with this property.

We now give correctness of levelled multiplication with the focal element
of \Cref{def:general-focal}.
\begin{proposition}[Levelled multiplication]
    Let $\ct,\ct'\in\calR_Q^2$ and $\pt\in\calR$.
    Suppose $\Dec(\ct)$ and $\pt$ encode $\bm z\in\bC^{N/2}$
    within error $(\epsilon_j)_{j\in[N/2]}$, while $\Dec(\ct')$ encodes
    $\bm z'\in\bC^{N/2}$ within error $(\epsilon'_j)_{j\in[N/2]}$,
    all at scale $\Delta$.
    For $\Mult^{[u]}(\ct,\ct')$ and $\PtMult^{[u]}(\pt,\ct')$,
    choose $\gamma$ according to
    \[
        \gamma\Delta\ge
        \begin{cases}
            2^b(E_\KS/\Delta+E_\RS),
                &\text{for }\Mult^{[u]},\\
            2^b E_\RS,
                &\text{for }\PtMult^{[u]}.
        \end{cases}
    \]
    In each case, the output decrypts to an encoding of $\bm z\odot\bm z'$
    at scale $\Delta$ within error $(\epsilon_{\times,j})_{j\in[N/2]}$, where
    \[
        \epsilon_{\times,j}
        =\abs{z_j}\epsilon'_j+\abs{z'_j}\epsilon_j
         +\epsilon_j\epsilon'_j+2^{-p_j}\gamma.
    \]
    Here, we require the size constraint
    \[
        \Delta\max_{j\in[N/2]}
        \bigl(\abs{z_jz'_j}+\epsilon_{\times,j}\bigr)<\frac{Q}{2\Delta}.
    \]
\end{proposition}
\begin{proof}
    Put $m_2=\Dec(\ct')$, and let $m_1=\Dec(\ct)$ for ciphertext--ciphertext
    multiplication or $m_1=\pt$ for plaintext--ciphertext multiplication.
    By \Cref{lem:ckks-ks-rs}, the output decrypts modulo $Q/\Delta$ to
    \[
        m_\times=\frac{m_1m_2}{\Delta}+ue,
        \qquad
        e=\begin{cases}
            e_\KS/\Delta+e_\RS,&\text{for }\Mult^{[u]},\\
            e_\RS,&\text{for }\PtMult^{[u]},
        \end{cases}
    \]
    where $\canorm{e_\KS}\le E_\KS$ and $\canorm{e_\RS}\le E_\RS$.
    The respective condition on $\gamma$ gives $2^b\canorm{e}\le\gamma\Delta$.
    For each $j\in[N/2]$, the input error bounds give
    \[
        \abs{\frac{\sigma_j(m_1m_2)}{\Delta^2}-z_jz'_j}
        \le\abs{z_j}\epsilon'_j+\abs{z'_j}\epsilon_j+\epsilon_j\epsilon'_j.
    \]
    Thus \Cref{lem:general-shape} gives the error bounds for $m_\times$.
    Finally,
    \[
        \norm{m_\times}_\infty\le\canorm{m_\times}
        \le\Delta\max_{j\in[N/2]}
        \bigl(\abs{z_jz'_j}+\epsilon_{\times,j}\bigr)<\frac{Q}{2\Delta}.
    \]
    Hence reduction modulo $Q/\Delta$ in decryption recovers $m_\times$, proving both claims.
\end{proof}

\subsubsection{Subring Focal Elements}
\label{sec:construction:subring-focal}

Fix a power of two $k \mid N/2$ and write $d=N/(2k)$.
Consider the subring $\calR'=\bZ[Y]/(Y^{2k}+1)\subseteq\calR$ with identification $Y=X^d$.
Choosing the focal element in this subring allows rotations within each coset
$r+k[d] \subset [N/2]$ without changing the error shape.

\begin{lemma}[Subring focal elements]
    \label{lem:subring-focal}
    Suppose $u\in\calR'$ has gains $(p_r)_{r\in[k]}$ and common $b$-bit overhead.
    Viewed as an element of $\calR$, it has $p_r$-bit gain on $r+k[d]$
    for each $r\in[k]$, with the same overhead.
    Furthermore, rotations $\kappa_{kt}$ fix $u$ and cyclically permute each coset $r+k[d]$.
\end{lemma}
\begin{proof}
    Under $Y=X^d$, evaluation at $\zeta^{5^j}$
    restricts to evaluation at $\zeta_{4k}^{5^j}$ on $\calR'$.
    Since $5$ has order $k$ modulo $4k$,
    $\sigma_j(u)=\sigma'_{j\bmod k}(u)$, where $\sigma'_r$
    denotes the $r$-th embedding of $\calR'$.
    Thus the embedding values repeat with period $k$,
    giving the claimed gains on each coset with the same overhead.

    Also, $5^{kt}\equiv1\pmod{4k}$, so $\kappa_{kt}$ fixes
    $\calR'$ pointwise and therefore fixes $u$.
    It is clear that $\kappa_{kt}$ corresponds to rotation by $k t$, so it rotates in each coset $r+k[d]$.
\end{proof}

\subsubsection{Focused Linear Map Evaluation}
\label{sec:concrete:circuits}

With a subring focal element as above, we can evaluate a separate linear map
on each coset $r+k[d]$, retaining its gain $p_r$.
The diagonal method of~\cite{HS18} evaluates these maps in parallel,
combining rotations by multiples of $k$ with plaintext multiplications.

\begin{lemma}[Linear map]
    \label{lem:linear-map}
    Suppose $u\in\calR'$ has gains $(p_r)_{r\in[k]}$ and common $b$-bit overhead.
    Given linear maps $A_r:\bC^d\to\bC^d$ for $r\in[k]$, define
    \[
        A=\sum_{r\in[k]}A_r\otimes\bm e_r\bm e_r^T:\bC^{N/2}\to\bC^{N/2}.
    \]
    so that it applies $A_r$ to the coset $r+k[d]$.
    Write $\norm{A_r}_{\infty\to\infty}=\max_i\sum_j|(A_r)_{i,j}|$.
    Suppose $\Dec(\ct)$ encodes $\bm z$ at scale $\Delta\ge2^bE_\Ecd$
    with error at most $\epsilon_r$ on $r+k[d]$.
    Choose $\gamma$ satisfying
    \[
        \gamma\Delta\ge C\,2^b \cdot d(E_\Ecd+E_\KS+E_\RS)
        \left(1+\norm{A}_{\infty\to\infty}+\norm{\bm z}_\infty
            +\max_{r\in[k]}\epsilon_r\right)
    \]
    for a sufficiently large constant $C$ independent of parameters.
    Then there is an operation
    \[
        \mathsf{Linear}^{[u]}_A:\calR_Q^2\to\calR_{Q/\Delta}^2
    \]
    such that $\Dec(\mathsf{Linear}^{[u]}_A(\ct))$ encodes $A\bm z$
    at scale $\Delta$ with error at most $\epsilon'_r$ on $r+k[d]$, where
    \[
        \epsilon'_r=\norm{A_r}_{\infty\to\infty}\epsilon_r
            +2^{-p_r}\gamma,
        \qquad \bm z_r=(z_{r+ki})_{i\in[d]}.
    \]
    Here, we require $\Delta\mid Q$ and
    \[
        \Delta\max_{r\in[k]}\{\norm{A_r\bm z_r}_\infty+\epsilon'_r\}
        <\frac{Q}{2\Delta}.
    \]
\end{lemma}
\begin{proof}
    For $t\in[d]$, define the diagonal vector $\bm a_t$ by
    \[
        (\bm a_t)_{r+ki}=(A_r)_{i,(i+t)\bmod d},
        \qquad r\in[k],\ i\in[d],
    \]
    and set $M_t=\Ecd^{[u]}(\bm a_t;\Delta)$.
    By \Cref{lem:general-encoding}, its encoding error on $r+k[d]$
    is at most $2^{-p_r} \cdot 2^bE_\Ecd/\Delta$.
    By \Cref{lem:subring-focal}, left rotation by $kt$, $\kappa_{kt}$, fixes $u$ and preserves each coset $r+k[d]$.
    We define
    \[
        \mathsf{Linear}^{[u]}_A(\ct)
        =\RS^{[u]}_\Delta\left(
            \sum_{t\in[d]}M_t \cdot \Aut^{[u]}_{\kappa_{kt}}(\ct)
        \right).
    \]
    
    Now, write $m=\Dec(\ct)$.
    Then, the output decrypts modulo $Q/\Delta$ to
    \[
        m'=\sum_t\frac{M_t}{\Delta}\kappa_{kt}(m)+ue',
        \qquad e'=\sum_t\frac{M_t}{\Delta}f_t+e_\RS,
    \]
    where $\canorm{f_t}\le E_\KS$ and $\canorm{e_\RS}\le E_\RS$.

    On a coset $r+k[d]$, the first sum applies $A_r$ ignoring the diagonal encoding error,
    giving error at most $\norm{A_r}_{\infty\to\infty}\epsilon_r$.
    Meanwhile, the diagonal encoding errors contribute at most
    \[
        2^{-p_r} \cdot \frac{2^b}{\Delta}
        d E_\Ecd(\norm{\bm z_r}_\infty+\epsilon_r)
    \]
    on the coset.
    Finally, since $\Delta\ge2^bE_\Ecd$ and $|(A_r)_{i,j}|\le\norm{A_r}_{\infty\to\infty}$,
    the diagonal encoding bounds give
    \begin{align*}
        \frac{\canorm{M_t}}{\Delta}
        &\le\norm{A}_{\infty\to\infty}+1,\\
        \canorm{e'}
        &\le dE_\KS(\norm{A}_{\infty\to\infty}+1)+E_\RS.
    \end{align*}
    By \Cref{lem:general-shape}, adding $ue'$ increases the encoding error
    on $r+k[d]$ by at most $2^{b-p_r}\canorm{e'}/\Delta$.
    Together with the diagonal encoding error, this is at most
    $2^{-p_r}\gamma$ by the condition on $\gamma$ and $d\ge1$,
    giving the desired error bound $\epsilon'_r$.

    The modulus bound ensures $\norm{m'}_\infty<Q/(2\Delta)$,
    so reduction modulo $Q/\Delta$ in decryption recovers $m'$.
    The construction uses one rescaling level, $d$ plaintext-ciphertext products,
    and at most $d-1$ rotations.
\end{proof}

\subsection{Constructions of Focal Element}
\label{sec:construction:focal}

We provide a concrete focal element construction through cyclotomic units.

\subsubsection{Cyclotomic units}
\label{sec:concrete:units}

Units in $\calR^\times$ provide attractive choices for focal elements,
since they form a multiplicative lattice, in which it is easier to search for desired precision.
As a bonus, all modulus are admissable for the units.

Following~\cite{cdpr2015}, we define the log-unit lattice as follows.
\begin{definition}[Log-unit lattice]
    Let \emph{logarithmic embedding} $\operatorname{Log} : \calR \to \bR^{N/2}$ as
    \[
        \operatorname{Log}(u)=(\log|\sigma_j(u)|)_{j\in[N/2]}.
    \]
    The \emph{log-unit lattice} is the lattice $\operatorname{Log}(\calR^\times) \subset \bR^{N/2}$.
\end{definition}

The log unit lattice lies on the hyperplane $V_0 = \{ x \in \bR^{N/2} : \sum_j x_j = 0 \}$,
since the field norm satisfies
\[
 \sum_{j\in[N/2]}\log|\sigma_j(u)|
 =\tfrac12\log|\operatorname{N}_{\calK/\bQ}(u)| = 0.
\]
By Dirichlet Unit Theorem, it is also a full-rank lattice in $V_0$.

Specifically, there is a useful sublattice given by cyclotomic units,
which admits a concrete representation.
The cyclotomic units
\[
    b_j = \frac{X^{5^j} - 1}{X - 1} \in \calR^\times, \qquad j \in [N/2] \setminus \{0\}.
\]
give a basis $\operatorname{Log}(b_j)$, which forms a full-rank sublattice $\operatorname{Log} C$.
We now construct focal units for arbitrary gain profiles.

\begin{proposition}[Construction of focal unit]
\label{prop:concrete-profile}
Given gains $p_j\ge0$ for $j\in[N/2]$,
denote its average $\bar p=\tfrac2N\sum_j p_j$.
There is a cyclotomic unit $u\in C$ with $p_j$-bit gain at each slot $j$
and common $(\bar p+\Gamma)$-bit overhead in the sense of
\Cref{def:general-focal}, where
\[
    \Gamma = \sup_{t \in V_0} \min_{v \in \operatorname{Log}C} \norm{t - v}_\infty
\]
is the $\ell_\infty$-covering radius of $\operatorname{Log}C$ in $V_0$.
\end{proposition}
\begin{proof}
    Set $t_j=\bar p-p_j$. Since $\sum_j t_j=0$, we have $\bm{t} = (t_j) \in V_0$.
    By the definition of $\Gamma$, there is a cyclotomic unit $u\in C$
    such that $\norm{\operatorname{Log}(u)-\bm t}_\infty\le\Gamma$.
    Therefore
    \[
        \log|\sigma_j(u)|\le t_j+\Gamma
        =(\bar p+\Gamma)-p_j,
        \qquad j\in[N/2],
    \]
    which gives the claimed gains and overhead.
\end{proof}

Thus every gain profile is achievable within $\Gamma$ of the overhead lower bound $b \ge \bar p$;
for a fixed ring, $\Gamma$ is independent of the gains.
In particular, taking gain $p$ on a fraction $f$ of slots and zero elsewhere
gives overhead $fp+\Gamma$, or $p/k+\Gamma$ when $f=1/k$.

\paragraph{In Practice.}
Focal units can be found through either rounding with respect to a reduced basis,
or mixed-integer linear programming.
Searching in a small subring reduces the number of variables
and gives the compatible rotations.

We formulate the integer program as follows.
For $k\ge2$, consider the real cyclotomic units in $\calR'$
\[
    c_j=Y^{-(j-1)/2}\frac{Y^j-1}{Y-1},
    \qquad j\in\{3,5,\ldots,2k-1\},
\]
and define $B_{i,j}=\log\abs{\sigma'_i(c_j)}$ for $i\in[k]$.
For target gains $p_i \ge 0$ on the subring slots $i\in[k]$,
we solve the mixed-integer linear program
\begin{equation}
    \label{eq:construction:focal-search}
    \begin{aligned}
        \text{minimize}\quad & b\\
        \text{subject to}\quad
        &(B\bm x)_i\le b-p_i,\qquad i\in[k],\\
        &\bm x\in\bZ^{k-1},\qquad b\in\bR.
    \end{aligned}
\end{equation}
This minimizes overhead subject to the specified gain profile.
The resulting unit $u=\prod_j c_j^{x_j}$ has logarithmic embedding $\bm v=B\bm x$,
and hence $p_i$-bit gain at slot $i$ with $b$-bit overhead.
The profile extends onto the full ring as in \Cref{lem:subring-focal}.

\subsection{Focused Slot Bootstrapping}
\label{sec:construction:focused-slot-bootstrapping}

Here, we present the focused slot bootstrapping construction from our earlier work~\cite{CL26}.

Consider the subring construction of \Cref{sec:construction:subring-focal},
specialized to a real focal unit $u \in \calR'$ with gains $(p,0,\ldots,0)$ and common $b$-bit overhead.
We suppose messages are supported on the focus slots $k[d]$, where $d=N/(2k)$,
and we assume $k\mid d$.
The focused encoding supports $p$-bit precision with scale dependent on overhead $b$,
rather than requiring the full $p$-bit scale.
Hence, focused computation can take smaller scale factor,
reducing the modulus consumption compared to the ordinary encoding.

Such focused encoding admits a concretely efficient bootstrapping,
which is close to a conventional bootstrapping except that \CtS{}, \StC{} are modified to preserve focus.
The slot bootstrapping proceeds in the order
\[
  \StC{}\longrightarrow\ModRaise{}\longrightarrow
  \CtS{}\longrightarrow\EvalMod{},
\]
starting and ending with messages in the focus slots.

There are two notable challenges for such a bootstrapping.
First, it has to put the message into coefficient encoding,
where the focused encoding does not apply.
Hence, we have to allocate scale factor of $p$-bit size to support $p$-bits of precision in coefficient encoding.
While this overhead is unavoidable in the slot bootstrapping,
we can minimize its impact on the modulus consumption by staying focused as long as possible,
only charging the full-size overhead at base modulus.

Another issue is comes with the packing.
The coefficient encoding is fully packed,
while the focused encoding can contain only $d = N/(2k)$ entries in the focus slots.
Hence, we represent this fully packed ciphertext using $k$ ciphertexts in focused encoding,
and show that the conversion maps can be computed efficiently.

\subsubsection{Interleaving and Focus Conversion}

Write $u_F\in(0,1)$ for the common value of $\sigma_{ki}(u)$, $i\in[d]$.
For a slot vector $\bm z$ supported on these slots, we have
\[
    \sigma(u)\odot\bm z=u_F\bm z.
\]
In particular, if $\ct\in\calR_Q^2$ satisfies
\[
    \Dec_\sk(\ct)\equiv\Delta\sigma^{-1}(\bm z)+ue\pmod Q,
\]
multiplying $u^{-1}$ gives
\[
    \Dec_\sk(u^{-1}\ct)
    \equiv\frac{\Delta}{u_F}\sigma^{-1}(\bm z)+e\pmod Q.
\]
Thus the same slot vector is represented at scale $\Delta/u_F$ with ordinary error $e$.
Conversely, multiplication by $u$ restores scale $\Delta$ and error shape $ue$.
Since $u$ and $u^{-1}$ are integral, these conversions are exact, and they consume no level.

After this conversion, we can pack the focus-slot messages of $k$ ciphertexts
at the common scale $\Delta/u_F$ using
\[
    \mathsf{Interleave}(\ct_0,\ldots,\ct_{k-1})
    = \sum_{j\in[k]}\mathsf{Rot}_{-j}(\ct_j).
\]
where $\Rot_j=\Aut_{\kappa_j}$ denotes left rotation by $j$.
For the $j$-th ciphertext, the rotation sends slots $k[d]$ to the coset $k[d]+j$.
The cosets partition all slots, so the sum packs the slot vectors into one ciphertext.

The reverse operation, $\mathsf{Deinterleave}$, needs to select the slots $k [d] + j$
and rotate it back into the coset $k[d]$ of the $j$-th ciphertext.
The required masking will be performed in \ModRaise{}, which eliminates a need for consuming modulus.

\subsubsection{Factorization of Coefficient--Slot Transforms}
\label{sec:construction:focused-transforms}

For $\bm m\in\bC^d$ with $m_i$ stored in focus slot $ki$, we choose
the target coefficient encoding with complex coefficient $m_i$ at index $ki$.
The conversion maps are
\begin{align*}
  \mathcal L_{\mathrm{S2C}}(\bm m)
    &=(\bm V_d\bm m\mid\cdots\mid\bm V_d\bm m),\\
  \mathcal L_{\mathrm{C2S}}(\bm z_0\mid\cdots\mid\bm z_{k-1})
    &=\frac1k\sum_{j\in[k]}\bm V_d^{-1}\bm z_j.
\end{align*}
Here, $\bm{V}_d$ is the DFT matrix given in \Cref{sec:prelim:dft-decomp}.
The first identity follows because the target polynomial is a polynomial
in $X^k$, whose slot values are repeated with period $d$.
For the second, averaging the $k$ blocks cancels the contributions
of coefficients whose indices are not divisible by $k$;
applying $\bm V_d^{-1}$ then recovers the remaining coefficients.
Note that modulus raising can introduce overflow in every
coefficient, so the \CtS{} map must accept all $dk$ slots.

The conversion maps connect sparse focus slots with a fully packed slot vector,
which calls for the interleaving map on the boundary.
To keep the scale governed by the overhead,
we seek a factorization in which only the interleaving step leaves focused encoding.
All preceding operations must therefore use operations preserving the focus,
like linear maps, rotations, and plaintext multiplications on the focus slots.

Now, we give the explicit decomposition.
For $\bm Z=[\bm z_0\mid\cdots\mid\bm z_{k-1}]\in\bC^{d\times k}$,
write $\mathcal J$ for interleaving its columns, so that
$(\mathcal J\bm Z)_{ki+j}=(\bm z_j)_i$.
We use the following linear maps.
\begin{enumerate}
  \item The reordered DFT $\mathcal T:\bC^d\to\bC^d$ applies a permutation
    to the output of $\bm V_d$, defined by
    \[
      (\mathcal T\bm m)_{j(d/k)+i}=(\bm V_d\bm m)_{ki+j},
      \qquad i\in[d/k],\ j\in[k].
    \]
  \item Let $\rho$ denote cyclic left rotation on $\bC^d$.
    The rotating fan-out $\mathcal O:\bC^d\to\bC^{d\times k}$
    and its adjoint are defined by
    \begin{align*}
      \mathcal O(\bm z)
        &=[\bm z\mid\rho^{-d/k}\bm z\mid\cdots\mid\rho^{-(k-1)d/k}\bm z],\\
      \mathcal O^*(\bm Z)&=\sum_{j\in[k]}\rho^{jd/k}\bm z_j.
    \end{align*}
  \item The block shuffle $\mathcal E:\bC^{d\times k}\to\bC^{d\times k}$
    permutes blocks across columns. Write
    \[
      \bm Z=
      \begin{bmatrix}
        \bm z_{0,0}&\cdots&\bm z_{0,k-1}\\
        \vdots&&\vdots\\
        \bm z_{k-1,0}&\cdots&\bm z_{k-1,k-1}
      \end{bmatrix},
      \qquad \bm z_{\ell,j}\in\bC^{d/k}.
    \]
    Then $\mathcal E(\bm Z)=\bm W$, whose blocks satisfy
    \[
      \bm w_{\ell,j}=\bm z_{\ell,(\ell-j)\bmod k},
      \qquad \ell,j\in[k].
    \]
\end{enumerate}

\begin{lemma}[Factorization of coefficient--slot transforms]
\label{lem:transform-factorization}
The conversion maps admit decompositions
\begin{equation}
  \label{eq:construction:transform-factorization}
  \mathcal L_{\mathrm{S2C}}=\mathcal J\mathcal E\mathcal O\mathcal T,
  \qquad
  \mathcal L_{\mathrm{C2S}}
    =\frac1k\mathcal T^{-1}\mathcal O^*\mathcal E\mathcal J^{-1}.
\end{equation}
\end{lemma}
\begin{proof}
To verify the first identity, write $\mathcal T\bm m$ as $k$ consecutive
blocks $(\bm v_0 \mid \cdots \mid \bm v_{k-1})$ of size $d/k$,
whose entries satisfy
\[
  (\bm v_j)_i=(\bm V_d\bm m)_{ki+j},
  \qquad i\in[d/k],\ j\in[k].
\]
The fan-out $\mathcal O$ forms column $j$ by rotating these blocks
to the right by $j$ positions,
so that block $\ell$ of column $j$ is $\bm v_{\ell-j}$, with block indices taken modulo $k$.
The shuffle $\mathcal E$ then retrieves block $\ell$ of output column $j$
from input column $\ell-j$, which contains $\bm v_{\ell-(\ell-j)}=\bm v_j$,
so every output column $j$ consists of $k$ copies of $\bm v_j$.

Then, interleaving $(\bm v_0,\ldots,\bm v_{k-1})$ restores the original
order of $\bm V_d\bm m$.
Since each block row contains these same blocks,
$\mathcal J$ produces $k$ consecutive copies of $\bm V_d\bm m$,
proving the first identity.

The second follows by taking adjoints, using $\mathcal E^*=\mathcal E$,
$\mathcal J^*=\mathcal J^{-1}$, and $\mathcal T^{-1}=\mathcal T^*/d$.
\end{proof}

In homomorphic evaluation, we represent each column with a ciphertext with the entries in the focus slots.
\Cref{alg:focused-transforms} implements the decomposition of \Cref{lem:transform-factorization}.
Here, $\mathsf{bl}_\ell$ is the focused encoding at scale $\Delta$ of the mask vector selecting $\ell d + k[d/k]$.
Note that since $k\mid d$, rotations by multiples of $d$ fix the element $u$, and the linear maps also preserve error shape.
\begin{algorithm}[htbp]
  \caption{Focused coefficient--slot transformations}
  \label{alg:focused-transforms}
  \begin{algorithmic}[1]
    \Procedure{\StC{}}{$\ct$}
      \State $\ct\gets\mathsf{Linear}^{[u]}_{\mathcal T}(\ct)$
      \State $(\ct_j)_{j\in[k]}\gets
        (\Aut^{[u]}_{\kappa_{-jd}}(\ct))_{j\in[k]}$ \Comment{fan-out}
      \For{$j\in[k]$}
        \State $\ct'_j\gets\RS^{[u]}_\Delta\left(\sum_{\ell\in[k]}\mathsf{bl}_\ell\ct_{\ell-j}\right)$ \Comment{block shuffle}
      \EndFor
      \State \Return $\mathsf{Interleave}(u^{-1}\ct'_0,\ldots,u^{-1}\ct'_{k-1})$
        \Comment{remove focus and interleave}
    \EndProcedure
    \Statex
    \Procedure{\CtS{}}{$\ct$}
      \State $(\ct_j)_{j\in[k]} \gets u \cdot \mathsf{Deinterleave}(\ct)$ \Comment{focus conversion}
      \For{$j\in[k]$}
        \State $\ct'_j\gets\RS^{[u]}_\Delta\left(\sum_{\ell\in[k]}\mathsf{bl}_\ell\ct_{\ell-j}\right)$ \Comment{block shuffle}
      \EndFor
      \State $\ct\gets\sum_{j\in[k]}\Aut^{[u]}_{\kappa_{jd}}(\ct'_j)$ \Comment{fan-in}
      \State \Return $\mathsf{Linear}^{[u]}_{\mathcal T^{-1}/k}(\ct)$
    \EndProcedure
  \end{algorithmic}
\end{algorithm}

We next present $\mathsf{Deinterleave}$ combined with \ModRaise{}
to avoid consuming a level for masking.

\subsubsection{Deinterleaving during Modulus Raising}

The $\mathsf{Deinterleave}$ needs masking to select slots.
Instead of removing the masking itself,
we eliminate the level cost of masking by applying it on \ModRaise{}-ed ciphertext.
We note that the \ModRaise{}-masking technique can be of independent interest.

Let $\hat\ct = (b,a) \in \calR^2$ be a lift of $\ct\in\calR_q^2$.
For a plaintext mask $M \in \RealExt$, rounding gives
\[
    \round{M b} + \round{M a} \cdot \sk
    = M (b + a \cdot \sk) + e, \qquad \canorm{e} \le E_\RS.
\]
Specifically, this operation does not cost any modulus, so we can pose aribtrary precision to the mask.

Using the technique,
$\mathsf{Deinterleave}$ can be defined by the mask $\mathsf{mask}_F \in \RealExt$ which selects the focus coset $k [d]$,
as follows.
\[
    \mathsf{Deinterleave}(\ct) = (\Rot_j(\ct'_j))_{j \in [k]},
    \qquad \ct'_j = [\, \round{\kappa_{-j}(\mathsf{mask}_F) \cdot \hat\ct} \,]_Q \in \calR_Q^2.
\]
Here, the $j$-th ciphertext selects the coset $k[d] + j$ via the mask, and rotate it into the focus coset $k[d]$.
After that, multiply by $u$ to enter the focused encoding for each ciphertext.

\subsubsection{Modular Reduction and Slot Bootstrapping}
We apply \EvalMod{} separately to the real and imaginary parts using focused multiplications.
For the modular-reduction inputs of \Cref{sec:prelim:ckks-bts},
let $G=q_0/\Delta'$ be the gap, where $\Delta'$ is the coefficient scale produced by \StC{}.
Han--Ki's sine interpolation~\cite[Section~4.1]{HK20}
yields modular-reduction approximation error $O(G^{-2})$ relative to the message,
so taking gap as $G=2^{n/2+O(1)}$ gives $n$-bit approximation accuracy.

Note that after modulus raising, \CtS{} and \EvalMod{} use scale $q_0=G\Delta'$,
a factor $G$ larger than the scale produced by \StC{}.
If we use the same unit for the processes, the scale for focused encoding would also grow by $G$.
To reduce this scale, we instantiate the preceding construction with two distinct units:
$u_{\rm S}$ of gain $p_{\rm S}$ at scale $\Delta_{\rm S}$ for \StC{},
and $u_{\rm C}$ of gain $p_{\rm C}$ at scale $\Delta_{\rm C}$ for \CtS{} and \EvalMod{}.
Write $u_{{\rm S},F}$ and $u_{{\rm C},F}$ for their respective values on the focus slots.
Focus conversion gives $\Delta_{\rm C}=u_{{\rm C},F} \cdot q_0$.

As in the standard bootstrapping, we interpret the \EvalMod{} output at scale $\Delta_{\rm C}/G$.
Multiplying by $u_{\rm S}u_{\rm C}^{-1}$ restores the original error shape
and scale $\Delta_{\rm S}$ exactly, without consuming a level.

\paragraph{Modulus cost of Slot Bootstrapping.}
Consider the ring degree, overflow, and other parameters fixed as $n$ grows.
For \StC{}, choose $p_{\rm S}=n+O(1)$ and $\Delta_{\rm S}=2^{n/k+O(1)}$.
The construction of \Cref{prop:concrete-profile} gives
$\Delta'=\Delta_{\rm S}/u_{{\rm S},F}=2^{n+O(1)}$,
so $q_0=G\Delta'$ requires $\tfrac32 n+O(1)$ bits.

For \CtS{} and \EvalMod{}, choose
$p_{\rm C}=n+\log G+O(1)=\tfrac32 n+O(1)$,
giving $n+O(1)$-bit accuracy relative to the message.
Thus the scale factors have bit lengths
\[
    \log\Delta_{\rm S}=\frac{n}{k}+O(1),
    \qquad
    \log\Delta_{\rm C}=\frac{3}{2} \frac{n}{k}+O(1).
\]
Let $D_{\StC{}}$, $D_{\CtS{}}$, and $D_{\EvalMod{}}$ denote the number of levels used by the respective stages.
The total bit size of bootstrapping modulus is therefore
\begin{align*}
    &\log q_0+D_{\StC{}}\log\Delta_{\rm S}
        +(D_{\CtS{}}+D_{\EvalMod{}})\log\Delta_{\rm C}\\
    &\qquad=\frac32n+\frac nk
        \left(D_{\StC{}}+\frac32(D_{\CtS{}}+D_{\EvalMod{}})\right)
        +O(1).
\end{align*}
Thus the base modulus contributes $\tfrac32n+O(1)$ bits once,
while the precision-dependent part of modulus consumption of all remaining levels
is reduced by a factor of $k$.

\paragraph{DFT decomposition for implementation.}
With a suitable ordering of the input focus slots, the reordered DFT is obtained
by conjugating the usual bit-reversed DFT by the reordering permutation.
This permutation rearranges the bits of the slot index, so each butterfly factor
retains its three-diagonal structure.
The usual layer-grouping tradeoff therefore applies to both transforms.

Each direction evaluates the DFT only once, despite temporarily using $k$ ciphertexts.
The additional work consists of $2(k-1)$ rotations, $O(k^2)$
plaintext--ciphertext products, and one rescaling level for the block shuffle.
The additional overhead remains modest for small $k$, making the construction suitable for practical parameter choices.

\paragraph{AI Disclosure.}
We used GPT-6 Astra to assist with manuscript preparation and, to a lesser extent, error
analysis and parameter selection.
The research ideas are all originated from the authors.
The authors have checked all technical details and take responsibility for the manuscript.

\phantomsection
\addcontentsline{toc}{section}{References}
\bibliographystyle{alpha}
\bibliography{common/refs}

\newcommand{\etalchar}[1]{$^{#1}$}
\begin{thebibliography}{MLWW26b}

\bibitem[Alb17]{Albrecht17}
Martin~R Albrecht.
\newblock On dual lattice attacks against small-secret lwe and parameter
  choices in helib and seal.
\newblock In {\em Annual International Conference on the Theory and
  Applications of Cryptographic Techniques}, pages 103--129. Springer, 2017.

\bibitem[BCC{\etalchar{+}}22]{BCCKK22}
Youngjin Bae, Jung~Hee Cheon, Wonhee Cho, Jaehyung Kim, and Taekyung Kim.
\newblock {META-BTS}: Bootstrapping precision beyond the limit.
\newblock In {\em ACM SIGSAC CCS}, 2022.

\bibitem[BCK{\etalchar{+}}23]{BCKPS23}
Youngjin Bae, Jung~Hee Cheon, Jaehyung Kim, Jai~Hyun Park, and Damien
  Stehl{\'e}.
\newblock {HERMES}: Efficient ring packing using {MLWE} ciphertexts and
  application to transciphering.
\newblock In {\em Advances in Cryptology -- {CRYPTO}}, pages 37--69, 2023.

\bibitem[BGV12]{BGV12}
Zvika Brakerski, Craig Gentry, and Vinod Vaikuntanathan.
\newblock (leveled) fully homomorphic encryption without bootstrapping.
\newblock In {\em Proceedings of the 3rd Innovations in Theoretical Computer
  Science Conference}, pages 309--325, 2012.

\bibitem[Bra12]{Bra12}
Zvika Brakerski.
\newblock Fully homomorphic encryption without modulus switching from classical
  {GapSVP}.
\newblock In {\em Advances in Cryptology -- {CRYPTO}}, pages 868--886, 2012.

\bibitem[BZ10]{BZ10}
Richard~P. Brent and Paul Zimmermann.
\newblock {\em Modern Computer Arithmetic}.
\newblock Cambridge University Press, 2010.

\bibitem[CCKS23]{CCKS23}
Jung~Hee Cheon, Wonhee Cho, Jaehyung Kim, and Damien Stehl{\'e}.
\newblock Homomorphic multiple precision multiplication for {CKKS} and reduced
  modulus consumption.
\newblock In {\em ACM SIGSAC CCS}, pages 696--710, 2023.

\bibitem[CCS19]{dft_decomp2}
Hao Chen, Ilaria Chillotti, and Yongsoo Song.
\newblock Improved bootstrapping for approximate homomorphic encryption.
\newblock In {\em Advances in Cryptology -- {EUROCRYPT}}, 2019.

\bibitem[CDPR16]{cdpr2015}
Ronald Cramer, L{\'e}o Ducas, Chris Peikert, and Oded Regev.
\newblock Recovering short generators of principal ideals in cyclotomic rings.
\newblock In {\em Annual International Conference on the Theory and
  Applications of Cryptographic Techniques}, pages 559--585. Springer, 2016.

\bibitem[CGGI20]{CGGI20}
Ilaria Chillotti, Nicolas Gama, Mariya Georgieva, and Malika Izabach{\`e}ne.
\newblock {TFHE}: Fast fully homomorphic encryption over the torus.
\newblock {\em Journal of Cryptology}, 33(1):34--91, 2020.

\bibitem[CHK{\etalchar{+}}18]{EC:CHKKS18}
Jung~Hee Cheon, Kyoohyung Han, Andrey Kim, Miran Kim, and Yongsoo Song.
\newblock Bootstrapping for approximate homomorphic encryption.
\newblock In {\em Advances in Cryptology -- {EUROCRYPT}}, 2018.

\bibitem[CKK20]{CKK20}
Jung~Hee Cheon, Dongwoo Kim, and Duhyeong Kim.
\newblock Efficient homomorphic comparison methods with optimal complexity.
\newblock In {\em Advances in Cryptology -- {ASIACRYPT}}, pages 221--256, 2020.

\bibitem[CKKS17]{ckks17}
Jung~Hee Cheon, Andrey Kim, Miran Kim, and Yongsoo Song.
\newblock Homomorphic encryption for arithmetic of approximate numbers.
\newblock In {\em International conference on the theory and application of
  cryptology and information security}, pages 409--437. Springer, 2017.

\bibitem[CKSS25]{CKSS25}
Hyeongmin Choe, Jaehyung Kim, Damien Stehl{\'e}, and Elias Suvanto.
\newblock Leveraging discrete {CKKS} to bootstrap in high precision.
\newblock In {\em ACM SIGSAC CCS}, 2025.

\bibitem[CL26]{CL26}
Jung~Hee Cheon and Junho Lee.
\newblock High-precision {CKKS} with focal units.
\newblock Cryptology {ePrint} Archive, Paper 2026/2531, 2026.

\bibitem[CLPX18]{CLPX18}
Hao Chen, Kim Laine, Rachel Player, and Yuhou Xia.
\newblock High-precision arithmetic in homomorphic encryption.
\newblock In {\em Topics in Cryptology -- {CT-RSA}}, 2018.

\bibitem[Dum26]{Dumezy26}
Jules Dumezy.
\newblock Fully homomorphic encryption for small batches of complex numbers.
\newblock Cryptology {ePrint} Archive, Paper 2026/2513, 2026.

\bibitem[FV12]{FV12}
Junfeng Fan and Frederik Vercauteren.
\newblock Somewhat practical fully homomorphic encryption.
\newblock Cryptology {ePrint} Archive, Paper 2012/144, 2012.

\bibitem[GV25]{GV25}
Robin Geelen and Frederik Vercauteren.
\newblock Fully homomorphic encryption for cyclotomic prime moduli.
\newblock In {\em Advances in Cryptology -- {EUROCRYPT}}, 2025.

\bibitem[GV26]{GV26BTS}
Robin Geelen and Frederik Vercauteren.
\newblock Better {GBFV} bootstrapping and faster encrypted edit distance
  computation.
\newblock {\em {IACR} Communications in Cryptology}, 2(4), 2026.

\bibitem[GZ26]{GZ26}
Mingyu Gao and Hongren Zheng.
\newblock {FHE} for {SIMD} arithmetic logic units with amortized {$O(1)$}
  bootstrapping per ciphertext.
\newblock In {\em Advances in Cryptology -- {CRYPTO}}, 2026.

\bibitem[HHC19]{dft_decomp}
Kyoohyung Han, Minki Hhan, and Jung~Hee Cheon.
\newblock Improved homomorphic discrete fourier transforms and {FHE}
  bootstrapping.
\newblock {\em IEEE Access}, 2019.

\bibitem[HK20]{HK20}
Kyoohyung Han and Dohyeong Ki.
\newblock Better bootstrapping for approximate homomorphic encryption.
\newblock In {\em Topics in Cryptology -- {CT-RSA}}, 2020.

\bibitem[HS18]{HS18}
Shai Halevi and Victor Shoup.
\newblock Faster homomorphic linear transformations in {HElib}.
\newblock In {\em Advances in Cryptology -- {CRYPTO}}, 2018.

\bibitem[HvdH21]{HV21}
David Harvey and Joris van~der Hoeven.
\newblock Integer multiplication in time {$O(n\log n)$}.
\newblock {\em Annals of Mathematics}, 193(2):563--617, 2021.

\bibitem[JL26]{JL26}
Daehyun Jang and Junho Lee.
\newblock Arithmetic for large-characteristic finite fields in {CKKS}.
\newblock Cryptology {ePrint} Archive, Paper 2026/1958, 2026.

\bibitem[Kim26]{Kim26GBFV}
Jaehyung Kim.
\newblock Bootstrapping {GBFV} with {CKKS}.
\newblock In {\em Advances in Cryptology -- {ASIACRYPT}}, 2026.
\newblock To appear; Cryptology ePrint Archive, Paper 2025/888.

\bibitem[LLFD26]{LLFD26}
Zhaoyang Liang, Runkun Liu, Jingwei Fang, and Dan Ding.
\newblock A new approach to reducing modulus requirements in homomorphic
  computation: Precision allocation in {CKKS}.
\newblock Cryptology {ePrint} Archive, Paper 2026/2562, 2026.

\bibitem[LLL{\etalchar{+}}21]{lee2021high}
Joon-Woo Lee, Eunsang Lee, Yongwoo Lee, Young-Sik Kim, and Jong-Seon No.
\newblock High-precision bootstrapping of {RNS-CKKS} homomorphic encryption
  using optimal minimax polynomial approximation and inverse sine function.
\newblock In {\em Advances in Cryptology -- {EUROCRYPT}}, 2021.

\bibitem[LMWW26]{LMWW26ALU}
Jiaming Liu, Shihe Ma, Anyu Wang, and Xiaoyun Wang.
\newblock High-precision homomorphic {ALU} over arbitrary moduli with {$O(1)$}
  bootstrapping.
\newblock Cryptology {ePrint} Archive, Paper 2026/1869, 2026.

\bibitem[LPR10]{LPR10}
Vadim Lyubashevsky, Chris Peikert, and Oded Regev.
\newblock On ideal lattices and learning with errors over rings.
\newblock In {\em Advances in Cryptology -- {EUROCRYPT}}, pages 1--23, 2010.

\bibitem[LW23]{LW23BatchII}
Feng-Hao Liu and Han Wang.
\newblock Batch bootstrapping {II}: Bootstrapping in polynomial modulus only
  requires {$\widetilde{O}(1)$} {FHE} multiplications in amortization.
\newblock In {\em Advances in Cryptology -- {EUROCRYPT}}, pages 353--384, 2023.

\bibitem[MLWW26a]{MLWW26BTS}
Shihe Ma, Jiaming Liu, Anyu Wang, and Xiaoyun Wang.
\newblock High-precision bootstrapping for {GCKKS}.
\newblock Cryptology {ePrint} Archive, Paper 2026/2600, 2026.

\bibitem[MLWW26b]{MLWW26}
Shihe Ma, Jiaming Liu, Anyu Wang, and Xiaoyun Wang.
\newblock Homomorphic encryption for high-precision arithmetic of approximate
  numbers.
\newblock Cryptology {ePrint} Archive, Paper 2026/2524, 2026.

\bibitem[WMGZ26]{WMGZ26}
Yuchen Wei, Kaisheng Ma, Mingyu Gao, and Hongren Zheng.
\newblock {FHE} for {ALU} over large prime moduli and application to one-round
  threshold {ECDSA}.
\newblock Cryptology {ePrint} Archive, Paper 2026/1836, 2026.

\end{thebibliography}

\end{document}